\documentclass[journal]{IEEEtran}

\usepackage{cite}
\usepackage{color,soul}
\usepackage{gensymb}
\usepackage{dsfont}
 \usepackage{bbm}

\ifCLASSINFOpdf
  \usepackage[pdftex]{graphicx}
  \graphicspath{{../pdf/}{../jpeg/}}
  \DeclareGraphicsExtensions{.pdf,.jpeg,.png}
\else
  \usepackage[dvips]{graphicx}
  \graphicspath{{../eps/}}
  \DeclareGraphicsExtensions{.eps}
\fi
\usepackage[cmex10]{amsmath}

\usepackage{amssymb}
{}

\usepackage{euscript}

\DeclareMathOperator*{\argmin}{arg\,min}
\usepackage{multirow}
\usepackage{algorithm}
\usepackage{algpseudocode}%[noend]
\usepackage{diagbox}
\usepackage{physics}
\usepackage{bm}
\usepackage{tikz}
\usepackage{comment}

\usepackage{amsmath,lipsum}
\usepackage{cuted}%%\stripsep-3pt

\usepackage{xcolor,colortbl}
\definecolor{myred}{RGB}{205 38 38}
 \usepackage{bbm}
\definecolor{myblue}{RGB}{105,89,205}
\definecolor{myorange}{RGB}{238,92,66}
\definecolor{mygray}{RGB}{205,201,201}

\ifCLASSOPTIONcompsoc
  \usepackage[caption=false,font=normalsize,labelfont=sf,textfont=sf]{subfig}
\else
  \usepackage[caption=false,font=footnotesize]{subfig}
\fi
\usepackage{url}
\allowdisplaybreaks[4]

\begin{document}
\title{Distributed Optimal Conservation Voltage Reduction in Integrated Primary-Secondary Distribution Systems}

\author{
         Qianzhi Zhang,~\IEEEmembership{Student Member,~IEEE,}
         Yifei Guo,~\IEEEmembership{Member,~IEEE,}
         Zhaoyu Wang,~\IEEEmembership{Senior Member,~IEEE,}
         
         and Fankun Bu,~\IEEEmembership{Student Member,~IEEE}
        % <-this % stops a space
\thanks{
%This work was supported in part by the U.S. Department of Energy Wind Energy Technologies Office under Grant DE-EE00008956, and in part by the National Science Foundation under Grant ECCS 1929975 (\emph{Corresponding author: Zhaoyu Wang}).
This work was supported in part by the U.S. Department of Energy Wind Energy Technologies Office under Grant DE-EE0008956, and in part by the National Science Foundation under ECCS 1929975 (\emph{Corresponding author: Zhaoyu Wang}).

The authors are with the Department of
Electrical and Computer Engineering, Iowa State University, Ames,
IA 50011 USA (e-mail: qianzhi@iastate.edu; yifeig@iastate.edu; wzy@iastate.edu; fbu@iastate.edu).}}

\maketitle

\begin{abstract}
This paper proposes an asychronous distributed leader-follower control method to achieve conservation voltage reduction (CVR) in three-phase unbalanced distribution systems by optimally scheduling smart inverters of distributed energy resources (DERs). One feature of the proposed method is to consider integrated primary-secondary distribution networks and voltage dependent loads. To ease the computational complexity introduced by the large number of secondary networks, we partition a system into distributed leader-follower control zones based on the network connectivity. To address the non-convexity from the nonlinear power flow and load models, a feedback-based linear approximation using instantaneous power and voltage measurements is proposed. This enables the online implementation of the proposed method to achieve fast tracking of system variations led by DERs. Another feature of the proposed method is the asynchronous implementations of the leader-follower controllers, which makes it compatible with non-uniform update rates and robust against communication delays and failures. Numerical tests are performed on a real distribution feeder in Midwest U. S. to validate the effectiveness and robustness of the proposed method.
\end{abstract}


\section*{Nomenclature}
\addcontentsline{toc}{section}{Nomenclature}
\begin{IEEEdescription}[\IEEEusemathlabelsep\IEEEsetlabelwidth{$V_1,V_2,V_3$}]
\item[\textbf{Sets and Indices}]
\item[$\mathcal{B}$] Set of boundary buses.
\item[$\mathcal{C}_{j}$] Set of children buses of bus $j$.
\item[$\mathcal{E}$] Set of branches.
\item[$\mathcal{N}$] Set of buses. $\mathcal{N}=\mathcal{P}\cup\mathcal{B}\cup\mathcal{S}$.
\item[$\mathcal{M}^t,\mathcal{N}^t$] Sets of follower controllers in asynchronous communication.
\item[$\mathcal{P}$] Set of primary network buses.
\item[$\mathcal{S}$] Set of secondary network buses.
\item[$\mathcal{X}$] Set of variables for primary network.
\item[$\mathcal{Z}_n$] Set of variables for secondary networks.
\item[$k$] Index of iteration.
\item[$n$] Index of secondary network.
\item[$t$] Index of time instant.
\item[$\phi$] Index of three-phase $\phi_a,\phi_b,\phi_c$.

\item[\textbf{Parameters}]
\item[$A_n,B_n$] Topology matrices for boundary system.
\item[$N_S$] Total number of secondary networks.
\item[$\widetilde{N}_S$] Setting number of secondary networks for partial barrier.
\item[$k_{1}^p,k_{2}^p,k_{3}^p$] Constant-impedance (Z), constant-current (I) and constant-power (P) coefficients for active ZIP loads.
\item[$k_{1}^q,k_{2}^q,k_{3}^q$] Constant-impedance (Z), constant-current (I) and constant-power (P) coefficients for reactive ZIP loads.
\item[$p^{\rm L}_{i,\phi,t}, q^{\rm L}_{i,\phi,t}$] Real and reactive load multipliers.
\item[$p^{g}_{i,\phi,t}$] Three-phase real power injections by the smart inverter.
\item[$q_{i,\phi,t}^{\rm cap}$] Three-phase reactive power capacity of smart inverters.
\item[$S^m_{ij,\phi,t}$] Three-phase apparent power measurements feedback from the system.
\item[$s^{\rm cap}_{i,\phi,t}$] Power capacity of smart inverters.
\item[$T$] Time length for termination.
\item[$v^{\rm min},v^{\rm max}$] Minimum and maximum limits for squared nodal voltage magnitude.
\item[$v^m_{ij,\phi,t}$] Three-phase voltage measurements feedback from the system.
\item[$z_{ij},r_{ij},x_{ij}$] Matrices of the line impedance, resistance and reactance.
\item[$\tau_n$] Setting iteration for boundary delay.
\item[$\mu,\tau^{\rm dec},\tau^{\rm inc}$] Parameters for updating penalty factor.

\item[\textbf{Variables}]
\item[$L_\rho$] Augmented Lagrangian.
\item[$P_{ij,\phi,t}$] Three-phase real power flows.
\item[$p_{i,\phi,t},q_{i,\phi,t}$] Three-phase active and reactive power bus injections.
\item[$p^{\rm ZIP}_{i,\phi,t}, q^{\rm ZIP}_{i,\phi,t}$] Three-phase real and reactive ZIP loads.
\item[$Q_{ij,\phi,t}$] Three-phase reactive power flows.
\item[$q^{g}_{i,\phi,t}$] Three-phase reactive power injections by the smart inverter.
\item[$r^k_n,s^k_n$] Primal and dual residuals.
\item[$S_{ij,\phi,t}$] Three-phase apparent power flow.
\item[$v_{i,\phi,t}$] Squared of three-phase voltage magnitude.
\item[$\bar{v}_{i,\phi,t}$] Estimation of the nonlinear term $\sqrt{v_{i,\phi,t}}$.
\item[$x,z_n$] Compact variables of primary and secondary networks.
\item[$x_{B,n},z_{B,n}$] Compact variables for boundary of primary network and coupling secondary network.
\item[$\lambda_n$] Lagrange multipliers.
\item[$\rho^k$] Iterative varying penalty coefficient for constraint violation.
\item[$\varepsilon_{ij,\phi}^{p},\varepsilon_{ij,\phi}^{q}$] Active and reactive power loss nonlinear terms.
\item[$\varepsilon_{i,\phi}^{v}$] Voltage drop nonlinear term.
\end{IEEEdescription}

% \fi
\IEEEpeerreviewmaketitle

\section{Introduction}
\IEEEPARstart{C}{onservation} voltage reduction (CVR) is to lower the voltage for peak load shaving and long-term energy savings, while maintaining the voltage at end users within the bound of set by American National Standards Institute (ANSI) \cite{ANSI,review_ZY}. 

Conventionally, CVR is implemented by rule-based or heuristic voltage controls at primary feeders by legacy regulating devices, such as on-load tap-changers, capacitor banks, step-voltage regulators, in slow timescales \cite{CVR_1,CVR_2}. The increasing integration of distributed energy resources (DERs), e.g., residential solar photovoltaics (PV), in secondary networks challenges conventional methods; but in turn, it also provides new voltage/var regulation capabilities by injecting or absorbing reactive power. The interactions between CVR and widespread DERs have been explored in \cite{ANL_VVC,QS_VVC,NREL_CVR_2}. It is demonstrated that DERs can flatten voltage profiles along feeders to allow deeper voltage reduction. In addition, the fast and flexible reactive power capabilities of four-quadrant smart inverters enable implementing CVR in fast timescales. To achieve system-wide optimal performance, voltage/var optimization based CVR (VVO-CVR), which can be cast into an optimal power flow  program, has spurred a substantial body of research. In \cite{CVR_VVO_2}, a linear least-squares problem is formulated for optimizing the CVR objective with a linearly approximated relation between voltages changes and actions of voltage regulating devices. The integration of optimal CVR and demand response is considered in \cite{CVR_VVO_4} to maximize the energy efficiency. Voltage optimization algorithm is developed in \cite{NREL_CVR} to implementing CVR by reactive power control of aggregated inverters. In \cite{VVO_FCS}, a convex optimization problem is formulated with network decomposition to optimally regulate voltages in a decentralized manner. In \cite{CVR_QZ}, the large-scale VVO-CVR problem is divided into a number of small-scale optimization problems using a distributed framework with only local information exchange, which coordinates multiple bus agents to obtain a solution for the original centralized problem. While the previous works have contributed valuable insights to VVO-CVR, there are problems remaining open, summarized as follows:

(1) \textit{Integrated Primary-Secondary Distribution Networks:} 
A practical distribution system is composed of medium-voltage (MV) primary networks and low-voltage (LV) secondary networks, where most loads and residential DERs are connected to secondary networks. However, previous studies have focused on primary networks while simplifying secondary network by using aggregate models to reduce computational burden. The grid-edge voltage regulation in distribution networks has not been well addressed. 

(2) \textit{Power Flow Models:} 
Some VVO-CVR studies have used full AC power flow models; however, the nonlinear nature makes the optimization programs non-convex and NP hard. Though heuristic algorithms (e.g. differential evolution algorithm \cite{zhou2017two}) or general nonlinear programming solvers (e.g. fmincon) can solve these problems, it often suffers the sub-optimality without proven optimal gaps. Other studies have directly dropped nonlinear terms (e.g. LinDistFlow) \cite{CVR_QZ} or used first-order Taylor expansion at a fixed point, to reduce the computational complexity \cite{ADMM_VVC_1}. However, such \emph{offline} linear approximation methods may bring non-negligible errors to power flow and bus voltage computation, thus, hindering the CVR performance. In addition, voltage-dependent load models must be used when studying CVR because the nature of CVR is that load is sensitive to voltage. Therefore, the nonlinear ZIP or exponential load models further complicate the VVO-CVR problem.

(3) \textit{Solution Algorithms:} 
The VVO-CVR can be directly solved by centralized solvers, which naturally requires global communication, monitoring, data collection and computation. Centralized solvers may be computationally expensive and less reliable for large systems, which is particularly true for a distribution system with a number of secondary networks. The  information privacy of customers is another concern for centralized control. To this end, some studies have developed distributed algorithms to solve VVO-CVR based on distribution optimization methods, such as alternating direction method of multipliers (ADMM) \cite{CVR_QZ} and primal-dual gradient algorithms \cite{Asy_VVC_2}. In \cite{CVR_QZ} and \cite{ADMM_VVC_7}, the ADMM is applied  to solve VVO-CVR in a three-phase unbalanced distribution network. In \cite{ADMM_VVC_1} and \cite{ADMM_VVC_6}, to provide a fully distributed solution, the convexified voltage regulation model is solved by ADMM. In \cite{ADMM_VVC_2}, different loading and PV penetration levels are tested for optimal reactive power control in large-scale distribution systems. In \cite{ADMM_VVC_3} and \cite{ADMM_VVC_4}, ADMM is implemented for solving the optimal reactive power dispatch problem of PV inverters. In \cite{ADMM_VVC_5}, optimal coordinated voltage control is achieved by ADMM for multiple distribution network clusters. Note that the distributed control algorithms in existing works inherently require synchronous update, which implies that the computation efficiency depends on the slowest agent. They are significantly affected by the differences in processing speed and communication delays, which may deteriorate the control performance \cite{Asy_VVC_1,Asy_VVC_3,Asy_VVC_4}. For example, the synchronous distributed algorithms may lose the fast-tracking capabilities for large systems.

To address these challenges, this paper proposes a leader-follower distributed algorithm based on asynchronous-ADMM (async-ADMM) \cite{asy_dis} to solve the VVO-CVR problem and enable online implementation with feedback-based linear approximation, where the primary network corresponds to the \emph{leader control} and each secondary network corresponds to a \emph{follower control}. The  contributions of this paper are threefold: 
\begin{itemize}
\item \textit{Mapping primary-secondary distribution system to ADMM-based leader-follower control framework:} To better model DERs' impacts and improve the grid-edge voltage regulation performance, we consider an integrated primary-secondary distribution system with detailed modeling of secondary networks. To solve the VVO-CVR problem in a distributed way, we first split the primary and secondary networks from modeling perspective, then introduce coupling constraints at boundary nodes, finally map the primary and secondary networks into leader and follower controllers in ADMM distributed framework.
\item \textit{Online feedback-based linear approximation method for power flow and ZIP load:} We propose an online feedback-based linear approximation method, where the instantaneous power and voltage measurements are used as system feedback in each iteration of ADMM to linearize the nonlinear terms of power flow calculation for both power flow and ZIP load models, which can significantly reduce the computational complexity and linearization errors by instantaneously tracking system variations.
\item \textit{Asynchronous implementation of ADMM:} We develop an asynchronous counterpart of conventional ADMM-based distributed control algorithms, which is robust against non-uniform update rates and communication delays, making it suitable for real-world applications.
\end{itemize}

The remainder of the paper is organized as follows: Section \ref{sec:framework} presents the overall framework of the proposed method. Section \ref{sec:cen_VVC} describes a centralized VVO-CVR in an integrated primary-secondary distribution system. Section \ref{sec:dis_VVC} proposes the distributed algorithm with online and asynchronous implementation. Simulation results and conclusions are given in Section \ref{sec:Results} and Section \ref{sec:Con}, respectively. 

\section{Overview of the Proposed Framework}\label{sec:framework}
\begin{figure}
	\vspace{-0pt} 
	\vspace{-0pt}
	\centering
	\includegraphics[width=1.0\linewidth]{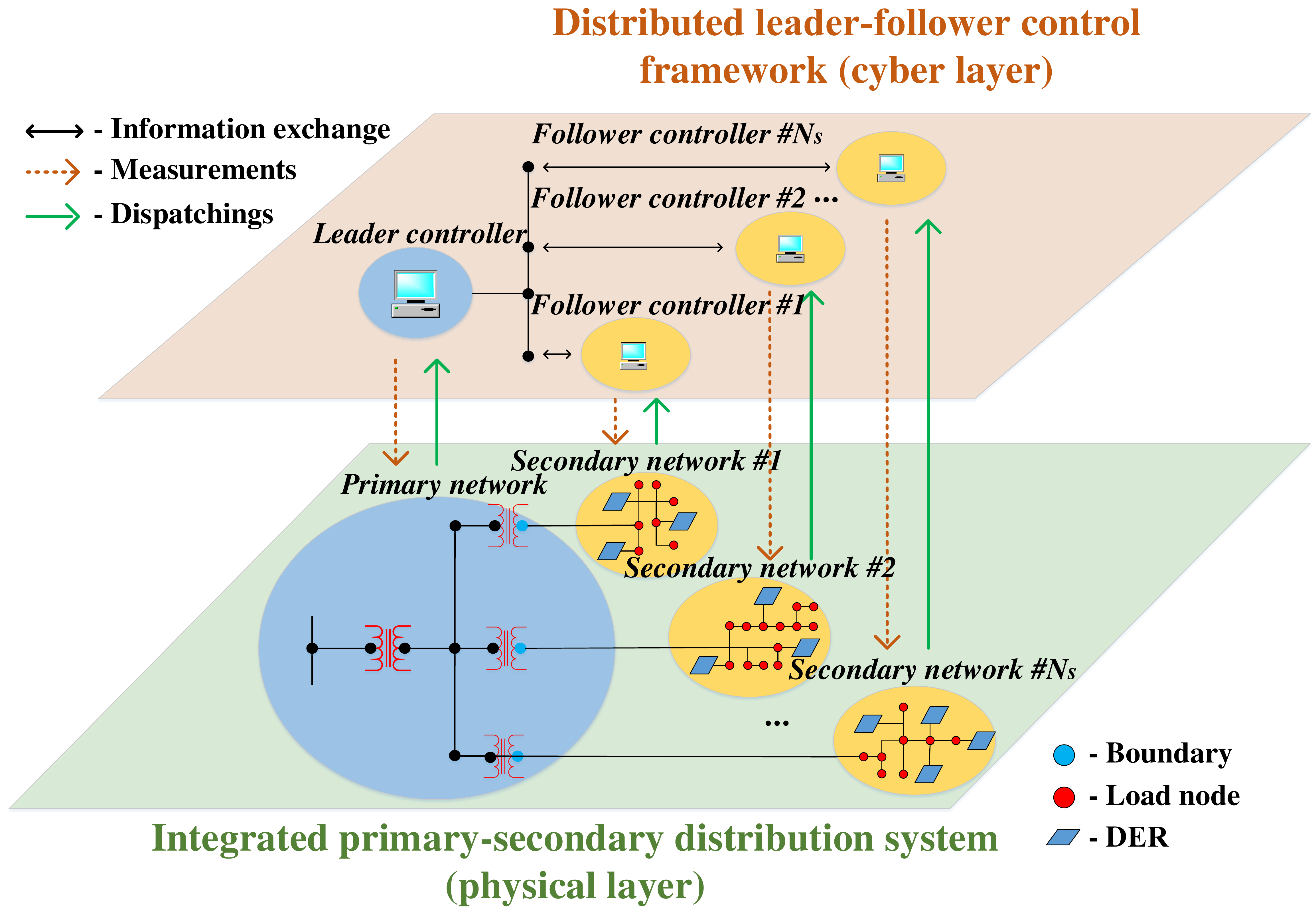}
	\vspace{-0pt} 
	\caption{Overall framework of the proposed distributed CVR with online and asynchronous implementations.}
	\centering
	\label{framework}
    \vspace{-0pt} 
\end{figure}

The general framework of the proposed distributed CVR with online and asynchronous implementations is shown in Fig. \ref{framework}. A VVO-CVR framework that dispatches smart inverters is developed for unbalanced three-phase distribution systems. The integration of primary-secondary networks with detailed secondary network models will be taken into account for better voltage regulation at grid-edge. Inspired by the physical structure of the distribution systems shown in Fig. \ref{framework}, the primary network corresponds to the leader controller and each secondary system corresponds to a follower controller. We then develop a distributed solution algorithm via ADMM framework to solve the VVO-CVR problem in a leader-follower distributed fashion, where the leader and followers controllers only exchange aggregate power and voltage magnitude information at boundaries. Note that, we specially address the asynchronous counterpart of the distributed solver to achieve robust and fast solutions while guaranteeing the convergence.

The nonlinear power flow and ZIP load models make the proposed problem nonconvex. To handle this issue, we propose to leverage voltage and line flow measurements as feedback to linearize these nonlinear models and make the program tractable. This feedback-based linear approximation method will be embedded within the distribution solution algorithm and combined with the online implementation of the distributed algorithm, where the reactive power outputs of smart inverters will be updated at each iteration by solving a time-varying convex optimization program in a leader-follower distributed fashion. In this way, we transform the conventional offline VVO-CVR to be an online feedback-based control model.

\section{Optimal CVR in Integrated Primary-Secondary Distribution Systems}\label{sec:cen_VVC}
\subsection{Modeling Integrated Primary-Secondary Distribution Networks}
A real distribution system consists of substation transformers, MV primary networks, service transformers, and LV secondary networks. Here, we consider a three-phase radial distribution system with $N$ buses denoted by set $\mathcal{N}$ and $N-1$ branches denoted by set $\mathcal{E}$. The buses in primary network and secondary networks are denoted by sets $\mathcal{P}$ and $\mathcal{S}$, respectively. The three-phase $\phi_a,\phi_b,\phi_c$ are simplified as $\phi$. The time instance is represented by $t$. For each bus $i\in\mathcal{N}$, $p^{\rm ZIP}_{i,\phi,t}, q^{\rm ZIP}_{i,\phi,t}\in\mathbbm{R}^{3\times1}$ are the vector of three-phase real and reactive ZIP loads at time $t$; $p^{g}_{i,\phi,t}, q^{g}_{i,\phi,t}\in\mathbbm{R}^{3\times1}$ are the vector of three-phase real and reactive power injections by the smart inverter at time $t$; $v_{i,\phi,t}:=V_{i,\phi,t}\odot V_{i,\phi,t}\in\mathbbm{R}^{3\times1}$ represents the vector of three-phase squared voltage magnitude at time $t$. $\mathcal{C}_{j}$ denotes the set of children buses. For any branch $({i,j})\in\mathcal{E}$, $z_{ij}=r_{ij}+{\bf i} x_{ij}\in\mathbbm{C}^{3\times3}$ are matrices of the three-phase branch resistance and reactance; $S_{ij,\phi,t}=P_{ij,\phi,t}+{\bf i} Q_{ij,\phi,t}\in\mathbbm{C}^{3\times1}$ denote the vector of three-phase real and reactive power flow from buses $i$ to $j$ at time $t$.

Most of the loads and DERs are connected to secondary networks, the power flows through the service transformers can be equivalently considered as the power injections $p_{i,\phi,t},q_{i,\phi,t}$ at the boundary bus $i\in\mathcal{B}$ (i.e., LV side bus of service transformer), where $\mathcal{B}\subseteq\mathcal{N}$ denotes the boundary bus set and let bus $i^\prime$ be the copy of bus $i$ at time $t$. Accordingly, the physical coupling of active power, reactive power and voltage at the boundary bus $i$ are expressed as,
\begin{align}
p_{i,\phi,t}+\sum_{j\in\mathcal{N}_i}P_{i^\prime j,\phi,t}&=0,\,\,\forall i\in\mathcal{B}\\
q_{i,\phi,t}+\sum_{j\in\mathcal{N}_i}Q_{i^\prime j,\phi,t}&=0,\,\,\forall i\in\mathcal{B}\\
v_{i,\phi,t}-{v_{i^\prime,\phi,t}}&=0,\,\,\forall i\in\mathcal{B}.
\end{align}

\subsection{VVO-Based CVR}
The aim of CVR is to reduce the total power consumption of the entire system while maintaining a feasible voltage profile across primary and secondary networks. Therefore, the VVO-CVR program can be formulated as follows,
\begin{subequations}
\begin{align}\label{eq_obj_CVR}
\text{min} \hspace{2mm}& \sum_{j:0\rightarrow j}\sum_{\phi\in\{\rm a,b,c\}}{\rm Re}\{S_{0j,\phi,t}\} \\
\nonumber\text{s.t.}\hspace{2mm}&\text{(1)-(3)}\\
P_{ij,\phi,t}&=\sum_{k:j\rightarrow k} P_{jk,\phi,t}-p^{g}_{j,\phi,t}+p_{j,\phi,t}^{\rm ZIP}+\varepsilon_{ij,\phi,t}^{p}\\
Q_{ij,\phi,t}&= \sum_{k:j\rightarrow k} Q_{jk,\phi,t}-q^{g}_{j,\phi,t}+q_{j,\phi,t}^{\rm ZIP}+\varepsilon_{ij,\phi,t}^{q}\\ 
v_{j,\phi,t}&=v_{i,\phi,t} - 2\big(\bar{r}_{ij}\odot P_{ij,\phi,t}+\bar{x}_{ij}\odot Q_{ij,\phi,t}\big) + \varepsilon_{i,\phi,t}^{v}\\
p^{\rm ZIP}_{i,\phi,t}&= p^{\rm L}_{i,\phi,t}\odot\big(k_{i,1}^p\cdot v_{i,\phi,t}+k_{i,2}^p\cdot\sqrt{v_{i,\phi,t}}+k_{i,3}^p\big)\\
q^{ZIP}_{i,\phi,t}&= q^{\rm L}_{i,\phi,t}\odot\big(k_{i,1}^q\cdot v_{i,\phi,t}+k_{i,2}^q\cdot \sqrt{v_{i,\phi,t}}+k_{i,3}^q\big)\\
v^{\rm min}&\leq v_{i,\phi,t}\leq v^{\rm max}, \forall i\in\mathcal{N}\\
-q_{i,\phi,t}^{\rm cap}&\leq q^g_{i,\phi,t}\leq q_{i,\phi,t}^{\rm cap}, \forall i\in\mathcal{G}.
\end{align}
\end{subequations}

In objective (4a), the ${\rm Re}\{S_{0j,\phi,t}\}$ denotes the three-phase active power supplied from the substation of the feeders at time $t$. For any branch $(i,j)\in\mathcal{E}$, the unbalanced three-phase branch flow model can be represented by constraints (4b)--(4d). Here, the $\odot$ and $\oslash$ denote the element-wise multiplication and division. If the network is not too severely unbalanced \cite{ADMM_VVC_1}, then the voltage magnitudes between the phases are similar and relative phase unbalance $\alpha$ is small. The unbalanced three-phase resistance matrix $\bar{r}_{ij}$ and reactance matrix $\bar{x}_{ij}$ can be referred to \cite{CVR_QZ}. The active and reactive ZIP loads $p^{\rm ZIP}_{i,\phi,t}$ and $q^{\rm ZIP}_{i,\phi,t}$ are calculated in constraints (4e) and (4f), where $p^{\rm L}_{i,\phi,t}, q^{\rm L}_{i,\phi,t}\in\mathbbm{R}^{3\times1}$ are the vectors of three-phase active and reactive load multipliers on bus $i$, respectively. $k_{i,1}^p$, $k_{i,2}^p$, $k_{i,3}^p$ and $k_{i,1}^q$, $k_{i,2}^q$, $k_{i,3}^q$ are constant-impedance (Z), constant-current (I) and constant-power (P) coefficients for active and reactive ZIP loads on bus $i$. Our work is proposing a distributed CVR model based on static optimal power flow problem, which focuses on system level optimization. The dynamic model, such as induction motor, is not included in the scope of our work. In constraint (4g), the (squared) bus voltage magnitude limits are set to the bus voltage $v^{\rm min}$ and $v^{\rm max}$, which are typically $[0.95^2,1.05^2]$ p.u., respectively. The nodal voltage constraint (4g) is applied to all nodes in the distribution system, including primary network and secondary networks.

In constraint (4h), the reactive power output of smart inverter is limited by the available reactive power of smart inverters $q_{i,\phi,t}^{\rm cap}$. Based on the capacity of the smart inverter $s^{\rm cap}_{i,\phi,t}$ and the active power output of smart inverter $p^g_{i,\phi,t}$, we can calculate the available capacity for reactive power generation of smart inverters $q_{i,\phi,t}^{\rm cap}$. According to the requirement for reactive power capability of the DERs in IEEE 1547-2018 Standard \cite{IEEE_1547}, the DERs shall provide voltage regulation capability by injecting reactive power or absorbing reactive power. Therefore, we assume there are enough reactive power capability for DER inverters in our proposed VVO-CVR problem. We also assume the DER system operates with the maximum power point tracking for active power control. Note that we focus on proposing an online distributed VVO-CVR to optimally dispatch the smart inverters in fast timescale. However, the conventional voltage regulation devices, such as on-load tap changer (OLTC) and capacitor banks (CBs), have slow reaction speed and limited number of switching operation, which cannot handle the fast changes in system states caused by loads and renewable energy resources in modern distribution systems. Thus, they should be controlled in a rather slow timescale instead of together with smart inverters, which is out of the scope of this paper. But it should be highlighted that, the operation of OLTC and CBs can be controlled by the leader controller, of which the impact can be taken into account in the fast timescale control of smart inverters. In this way, the coordination among them can be easily achieved. 

The power flow model (4b)--(4d) includes non-linear terms  $\varepsilon_{ij,\phi}^{p},\varepsilon_{ij,\phi}^{q}$ and $\varepsilon_{i,\phi}^{v}$. In the unbalanced three-phase branch flow model, these nonlinear terms render the program non-convex that is hard to solver. However, simply dropping these nonlinear terms may cause non-negligible modeling errors that deteriorates the voltage regulation performance. Similarly, when calculating active/reactive ZIP loads in constraints (4e) and (4f), the nonlinear part $\sqrt{v_{i,\phi,t}}$ also introduces non-convexity. To make the problem tractable, we propose to estimate the nonlinear terms with instantaneous voltage and line flow measurements, which can be referred to as a \emph{feedback-based linear approximation} method. Such approximate models of power flow and ZIP load are integrated with the online implementation of the distributed solver, which will be detailed in Section IV-C.

\subsection{Reformulating VVO-CVR for Distributed Solution by Splitting Primary and Secondary Networks}
We first compactly define the decision vector $x:=\left[p_{i,\phi,t},q_{i,\phi,t},v_{i,\phi,t}\right]^T,i\in\mathcal{P}$ for primary network and $z_n:=\left[P_{i^\prime j,\phi,t},Q_{i^\prime j,\phi,t},v_{i^\prime,\phi,t}\right]^T,i\in\mathcal{S}$ for $n$th secondary network, that consist of all the active/reactive branch flows and squared bus voltage magnitudes belonging to the primary network and $n$th secondary network, respectively. Accordingly, the boundary variables $x_{B,n}$ and $z_{B,n}$ (sub-vectors of $x$ and $z_n$, respectively) regarding $n$th secondary network (suppose bus $i$ is the boundary bus) can be compactly represented by: $x_{B,n}:=\left[p_{i,\phi,t},q_{i,\phi,t},v_{i,\phi,t}\right]^T,i\in\mathcal{B}$ and $z_{B,n}:=\left[\sum_{j\in\mathcal{C}_i}P_{i^\prime j,\phi,t},\sum_{j\in \mathcal{C}_i}Q_{i^\prime j,\phi,t},v_{i^\prime,\phi,t}\right]^T,i\in\mathcal{B}$, respectively. By decomposing the constraints into primary network, secondary networks and boundary systems, the VVO-CVR problem in (1)--(3) and (4) can be compactly reformulated as,
\begin{subequations}
\begin{align}\label{eq_compact_CVR}
\underset{x,z_n,\forall n}{\text{min}}\hspace{2mm}& f(x)\\
\text{s.t.}\hspace{2mm}&x\in\mathcal{X}:=\left\{x|\text{(4b)--(4g)}\right\}\\ 
&z_n\in\mathcal{Z}_n:=\left\{z_n|\text{(4b)--(4h)}\right\},\,\forall n\\ &{A}_nx_{B,n}+{B}_nz_{B,n}=0\iff\left\{\text{(1)--(3)}\right\},\forall n
\end{align}
\end{subequations}
where constraint sets (5d) is defined for boundary system. The $A_n=I_9$ and $B_n={\rm blkdiag}(I_6,-I_3)$ for three-phase secondary networks and $A_n=I_3$ and $B_n={\rm blkdiag}(I_2,-I_1)$ for single-phase secondary networks, where $I_{m}$ denotes the $m\times m$ identity matrix. 

\section{Proposed Distributed Solution Algorithm for Asynchronous and Online Implementations}\label{sec:dis_VVC}
\subsection{Standard Distributed Solution Algorithm via ADMM}
The augmented Lagrangian of the compact VVO-based CVR (5) is shown as, 
\begin{multline}\label{eq3_Lag}
L_\rho=f(x)+\sum_{n=1}^{N_S}\lambda_n\odot({A}_n\odot x_{B,n}+{B}_n\odot z_{B,n})\\
+\sum_{n=1}^{N_S}\frac{\rho^k}{2}\left\|{A}_n\odot x_{B,n}+{B}_n\odot z_{B,n}\right\|^2_{2}
\end{multline}
where the $\lambda_n$ is the vector of the Lagrange multipliers for the primary network (leader controller) and the coupling $n$th secondary network (follower controller), $k$ denotes the iteration index, and $\rho^k>0$ is the iterative varying penalty coefficient for constraint violation. 
 
The ADMM solves the problem (5) by alternatingly minimizing the augmented Lagrangian \eqref{eq3_Lag} over $x,z_n$ and $\lambda_n$. It consists of the following steps:
(i) By \eqref{ADMM_1}, the leader controller first updates the variables $x$ associated with primary system, where the update boundary variables $x^{k+1}_{B,n}$ will be sent to each corresponding follower controller. (ii) By \eqref{ADMM_2}, the follower controllers update the  variables $z_n$ associated with each secondary system by. Since each distributed follower controller only solves the problem in terms of the local variables in secondary systems so that this step can be performed in parallel. The updated boundary variables $z^{k+1}_{B,n}$ will be sent to the leader controller. (iii) As in \eqref{ADMM_3}, each follower controller is also responsible for updating the variables $\lambda_n$ by $x^{k+1}_{B,n}$ and $z^{k+1}_{B,n}$. The newly updated variables $\lambda^{k+1}_{n}$ will be sent to the leader controller. 
\begin{align}\label{ADMM_1}
\nonumber x^{k+1}\leftarrow&\argmin_{x\in\mathcal{X}} f(x)+\sum_{n=1}^{N_s}\lambda^k_n\odot({A}_n\odot x_{B,n}+{B}_n\odot z^k_{B,n})\\
&+\sum_{n=1}^{N_s}\frac{\rho^k}{2}\left\|{A}_n\odot x_{B,n}+{B}_n\odot z^k_{B,n}\right\|^2_{2},
\end{align} 
\begin{align}\label{ADMM_2}
\nonumber z^{k+1}_{n}\leftarrow&\argmin_{z_n\in\mathcal{Z}_n}\lambda^k_n\odot\big({A}_n\odot x^{k+1}_{B,n}+{B}_n\odot z_{B,n}\big)\\
&+\frac{\rho^k}{2}\left\|{A}_n\odot x^{k+1}_{B,n}+{B}_n\odot z_{B,n}\right\|^2_{2},
\end{align} 
\begin{align}\label{ADMM_3}
\lambda^{k+1}_{n}&\leftarrow\lambda^k_{n}+\rho^k({A}_n\odot x^{k+1}_{B,n}+{B}_n\odot z^{k+1}_{B,n}),
\end{align} 
where the sync-ADMM necessitates the use of a global clock $k$ for both leader controller and follower controllers. The convergence and optimality analyses of this conventional sync-ADMM can be found in \cite{boyd2011distributed}. 

\subsection{Asynchronous Implementation}
\begin{figure}
	\vspace{-0pt} 
	\vspace{-0pt}
	\centering
	\includegraphics[width=1.0\linewidth]{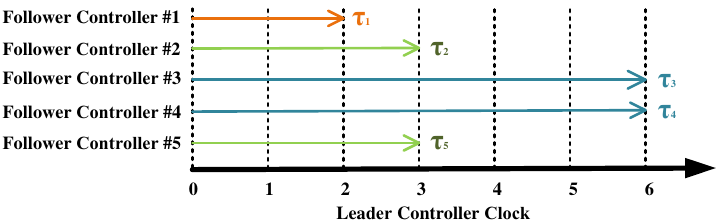}
	\vspace{-0pt} 
	\caption{An example of leader-follower async-ADMM framework.}
	\centering
	\label{example_async_admm}
   \vspace{-0pt} 
\end{figure}

When implementing sync-ADMM to solve the VVO-CVR in above formulations \eqref{ADMM_1}--\eqref{ADMM_3}, the leader controller of the primary network has to wait till all the follower controllers of the secondary networks finish updating their variables $z_n$ to receive the latest boundary variables $z_{B,n}$ and proceed. Thus, the sync-ADMM is not ideal for optimally dispatching smart inverters in a fast timescale and robust for communication delay. To alleviate this problem, an async-ADMM method \cite{asy_dis} is implemented, where the leader controller only needs to receive the updates from a minimum number of $\widetilde{N}_S\geq1$ follower controllers, and $\widetilde{N}_S$ can be much smaller than the total number of follower controllers $N_S$. This relaxation is the so called \emph{partial barrier}. Here a small number of $\widetilde{N}_S$ based on partial barrier means that the update frequencies of the slow follower controllers can be much less than those faster follower controllers. To ensure sufficient freshness of all the updates, we also require a \emph{bounded delay}, i.e., the $n$-th follower controller must communicate with the leader controller and receive the results from the leader controller for updating local variables at least once every $\tau_n\geq1$ iterations. Consequently, the update in every follower controller can be at most $\tau_n$ iterations later than the leader's clock. An example of the asynchronous update is given in Fig. \ref{example_async_admm}, where the partial barrier $\widetilde{N}_S=2$. In this example, the leader controller receives the updates from follower controller 1 at clock time two; the leader controller receives the updates from follower controllers 2 and 5 at clock time three; the leader controller receives the updates from follower controllers 3 and 4 at clock time six. Meanwhile, the leader controller has already preserved the update of follower controller 1 for five iterations and follower controllers 2 and 5 for four iterations.  

The convergence rate of this async-ADMM is in the order of $O({N_S\tau_n}/{2T\widetilde{N}_S})$ \cite{asy_dis}. The $T$ is the total time length for termination. This convergence rate can be intuitively explained by different value of $N_S$, $\widetilde{N}_S$ and $\tau_n$: (i) If the number of secondary networks in the system, $N_S$, is large, more iterations $k$ in the async-ADMM are needed for convergence. It is because each follower controller's update is less informative with a smaller data subset. (ii) If there is a large number $\widetilde{N}_S$ of secondary networks exchanging information with the primary network in the async-ADMM, the number of iterations $k$ required for convergence is reduced. This is because the primary network can collect more information from the secondary networks in each iteration. (iii) If a large $\tau_n$ exists, due to the very infrequent information exchange between the leader controller and follower controllers, a larger number of iteration $k$ is needed for convergence. To further improve the convergence performance and capture fast system variation of the async-ADMM, as well as make the performance less dependent on the initial choice, we implement an iterative varying penalty update \cite{boyd2011distributed} as follows,
\begin{align}\label{rho_1}  
    \rho^{k+1}:=\left\{\begin{matrix}\tau^{\rm inc}\rho^k,\,\,\,\text{if}\,\,\|r^k\|_2>\mu\|s^k\|_2\\{\rho^k}/{\tau^{\rm dec}},\,\,\,\text{if}\,\,\|s^k\|_2>\mu\|r^k\|_2\\\rho^k,\,\,\,\text{otherwise}\end{matrix}\right.
\end{align}
where $\mu>1$, $\tau^{\rm dec}>1$ and $\tau^{\rm inc}>1$ are the updating parameters. The primal and dual residuals $r^k_n$ and $s^k_n$ are calculated as,
\begin{align}\label{rho_2}  
r^k_n&={A}_n\odot x^k_{B,n}+{B}_n\odot z^k_{B,n},\forall n\\
s^k_n&=\rho_k{A}_n^T\odot {B}_n\left(z^{k+1}_{B,n}-z^k_{B,n}\right),\forall n.
\end{align}

\subsection{Online Implementation}
\begin{algorithm}[t]
\caption{Online and Asynchronous Implementations of Distributed VVO-CVR}\label{alg:AsynADMM}
\begin{algorithmic}[1]
\State \hspace{0mm}{\bf Initialization}: Set $t=0$ and choose $x(0), z_n(0), n=1,\ldots,N_S$.
\Repeat
\State $t\leftarrow t+1$.
\State If leader controller receives the newly updated  ${z}_{B,n}$ and $\lambda_n$ from some follower controller $n$,
then ${\mathcal{M}}^t\leftarrow{\mathcal{M}}^{t-1}\cup\{n\}$.
\State Let $\widetilde{z}^t_{B,n}\leftarrow z^t_{B,n},\widetilde{\lambda}^t_n\leftarrow \lambda^t_n,\,n\in{\mathcal{M}}^t$
and $\widetilde{z}^t_{B,n}\leftarrow\widetilde{z}^{t-1}_{B,n},\widetilde{\lambda}^t_n\leftarrow\widetilde{\lambda}^{t-1}_n, \,n\notin{\mathcal{M}}^t$.
  \If{$|{\mathcal{M}}^t|\geq\widetilde{N}_S$}
  \State Update $x^{t+1}$ by \eqref{ADMM_1} using $\widetilde{z}^t_{B,n}$.
  \State Send  $x^{t+1}_{B,n}$ to follower controller $n\in\mathcal{M}^t$.
  \State Reset $\mathcal{M}^t\leftarrow\emptyset$.
  \EndIf
\For{every $n\in\mathcal{N}^t$}
\State Update $z^{t+1}_{n}$ by \eqref{ADMM_2}.
\State Update $\lambda^{t+1}_{n}$ by \eqref{ADMM_3}.
\State Send $z^{t+1}_{B,n}$ and $\lambda^{t+1}_{n}$ to leader controller.
\EndFor
\For{every $n\notin\mathcal{N}^t$}
\State Let $z^{t+1}_n\leftarrow z^t_n$ and $\lambda^{t+1}_n\leftarrow\lambda^{t}_n$.
\EndFor
\State Update $\rho^t$ by (10)--(12).
\State Update reactive power output of inverters as per $z_n^{t+1}$.
\State Update the nonlinear terms $\varepsilon_{ij,\phi,t}^{p}, \varepsilon_{ij,\phi,t}^{q}$ and $\varepsilon_{i,\phi,t}^{v}$ by (13)--(15) with measurements feedback from the system.
\State Update the estimation of the nonlinear term $\bar{v}_{i,\phi,t}$ in ZIP loads (16)--(18) with measurements feedback from the system.
\Until $t$ terminates.
\end{algorithmic}
\end{algorithm}

To accurately track the fast variations of renewable generation and load demand for better CVR performance, we address the online implementation of the proposed distributed algorithm. In this context, we directly represent the iteration index by a symbol $t$ in the distributed algorithm. Specifically, the instantaneous power and voltage measurements at time $t-1$ are used as the system feedback to estimate the nonlinear terms of power flow and ZIP load models at time $t$. In this paper, we assume a widespread coverage of meters throughout the network. The leader and follower controllers have access to the instantaneous measurements of line flow and voltage. \footnote{If line flow measurements are not available, one can approximately estimate them through the linearized power flow model.} Thus, the nonlinear terms $\varepsilon_{ij,\phi,t}^{p}, \varepsilon_{ij,\phi,t}^{q}$ and $\varepsilon_{i,\phi,t}^{v}$ in 
(4b)--(4d) at time $t$ can be estimated as constants with the system feedback measurements from previous time $t-1$ as, 
\begin{align}
\varepsilon_{ij,\phi,t}^{p}=&{\rm Re}\{(S^m_{ij,\phi,t-1}\oslash v^m_{i,\phi,t-1})\odot(v^m_{i,\phi,t-1}-v^m_{j,\phi,t-1})\},\\
\varepsilon_{ij,\phi,t}^{q}=&{\rm Im}\{(S^m_{ij,\phi,t-1}\oslash v^m_{i,\phi,t-1})\odot(v^m_{i,\phi,t-1}-v^m_{j,\phi,t-1})\},\\
\nonumber\varepsilon_{i,\phi,t}^{v}=&\big[z_{ij}((S^m_{ij,\phi,t-1})^*\oslash (v^m_{i,\phi,t-1})^*)\big]\\
&\odot \big[z_{ij}^*(S^m_{ij,\phi,t-1}\oslash v^m_{i,\phi,t-1})\big],
\end{align}
where the $S^m_{ij,\phi,t-1}\in\mathbbm{R}^{3\times1}$, $v^m_{i,\phi,t-1}\in\mathbbm{R}^{3\times1}$ and $v^m_{j,\phi,t-1}\in\mathbbm{R}^{3\times1}$ are the instantaneous three-phase apparent power and voltage measurements feedback from the system at time $t-1$. Similarly, to handle the non-convexity due to the nonlinear part $\sqrt{v_{i,\phi,t}}$ in active/reactive ZIP loads, we use the first-order Talyor expansion to linearize it around the instantaneous voltage measurements $v^m_{i,\phi,t-1}$ as,
\begin{align}\label{ZIP_P_2} 
\nonumber \bar{v}_{i,\phi,t}=&v^m_{i,\phi,t-1}\\
&+\frac{1}{2}(v^m_{i,\phi,t-1})^{-1}\odot(v_{i,\phi,t}-v^m_{i,\phi,t-1}\odot v^m_{i,\phi,t-1}),
\end{align}
where $\bar{v}_{i,\phi,t}\in\mathbbm{R}^{3\times1}$ is the estimation of the nonlinear term $\sqrt{v_{i,\phi,t}}$. Therefore, the active and reactive ZIP loads in (4e) and (4f) are re-written as follows,
\begin{align}\label{ZIP_P_3} 
p^{\rm ZIP}_{i,\phi,t}&\simeq p^{\rm L}_{i,\phi,t}\odot(k_{i,1}^p\cdot v_{i,\phi,t}+k_{i,2}^p\cdot\bar{v}_{i,\phi,t}+k_{i,3}^p),\\
q^{\rm ZIP}_{i,\phi,t}&\simeq q^{\rm L}_{i,\phi,t}\odot(k_{i,1}^q\cdot v_{i,\phi,t}+k_{i,2}^q\cdot\bar{v}_{i,\phi,t}+k_{i,3}^q).
\end{align}

In this way, the above feedback-based linear approximation method with online system measurements can make the sub-problems of leader and follower controllers convex and can be efficiently solved. Due to the distributed solution algorithm, the original large-scale centralized VVO-CVR problem is decomposed to several sub-problems for leader controller of primary network and follower controllers of secondary networks, implying better a scalability. This is exactly an inherent advantage of distributed optimization techniques. The detailed procedure of the online async-ADMM is shown in Algorithm \ref{alg:AsynADMM}. The $\mathcal{M}^t$ denotes the set of follower controllers whose local updates have arrived at leader controller at iteration $t$ and $\mathcal{N}^t$ denotes set of follower controllers that receives the newly updated $x_{B,n}$ at iteration $t$. During the iteration, if the $n$th follower controller $n\notin\mathcal{N}^t$, which does not update the variable at iteration $t$, then the values of ${x}_{B,n}$, ${z}_{B,n}$ and ${\lambda}_{n}$ and ${x}_{B,n}$ remain unchanged until the newly updated values come. 

\section{Case Studies}\label{sec:Results}
\subsection{Simulation Setup}
\begin{figure}
	\vspace{-0pt} 
	\vspace{-0pt}
	\centering
	\includegraphics[width=0.8\linewidth]{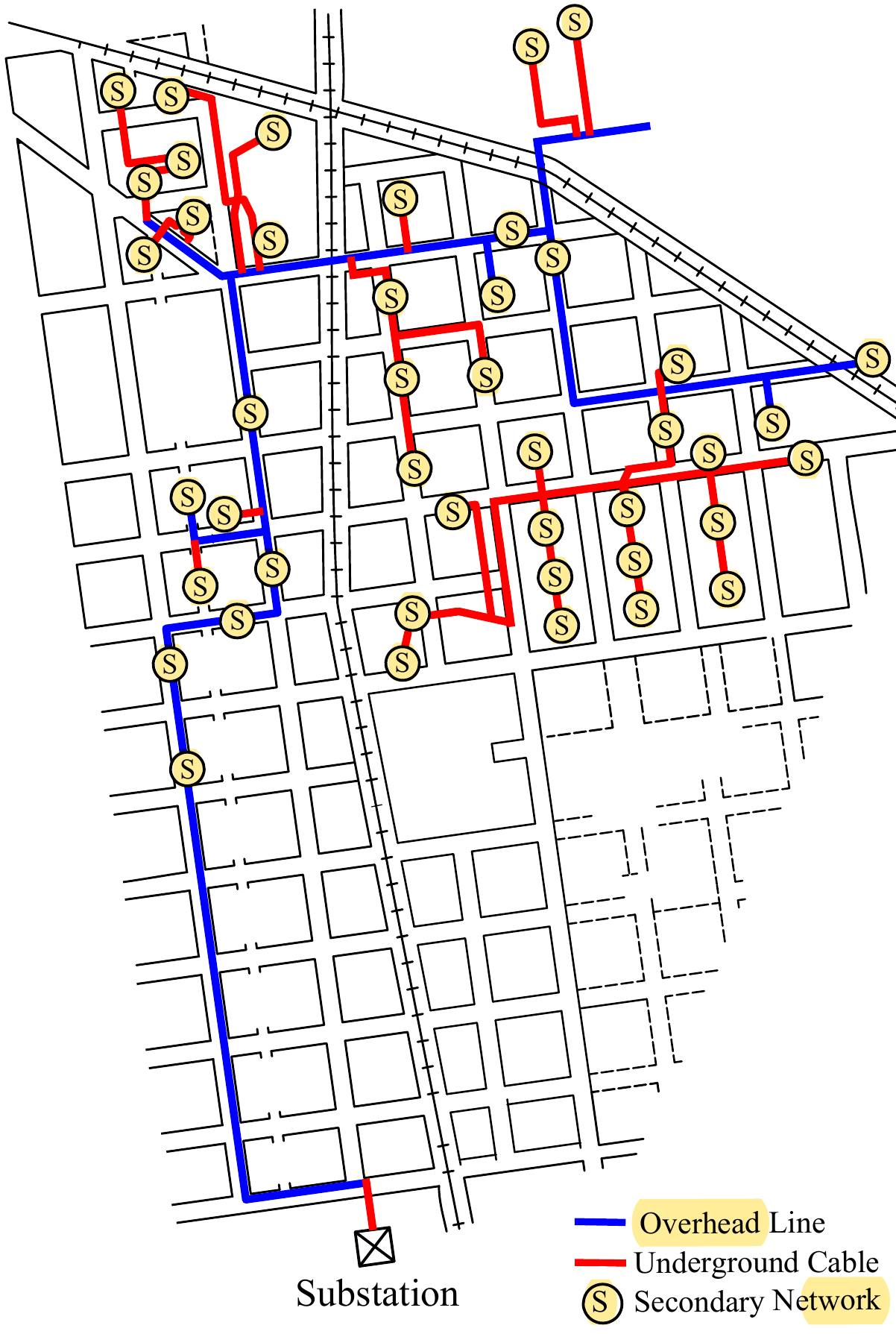}
	\vspace{-0pt} 
	\caption{A real primary-secondary distribution feeder in Midwest U.S. \cite{Realsystem}, consisting one MV primary network and forty-four LV secondary networks.}
	\centering
	\label{realsystem}
    \vspace{-0pt} 
\end{figure}

A real-world distribution feeder located in Midwest U.S. \cite{Realsystem} in Fig. \ref{realsystem} is used to illustrate our proposed scheme. This real feeder is shared by our utility partner, which consists of one primary network and forty-four secondary networks. The primary network is denoted by overhead lines (blue) and underground lines (red), and the secondary network is denoted by a circled capital letter S. Each secondary network includes a service transformer, a secondary circuit with multiple customers and DERs. We have two reasons for choosing this real distribution feeder as the test system: (i) The real distribution grid model \cite{Realsystem} is an integrated primary-secondary distribution, which can be used to verify our proposed distributed CVR model. While most of the IEEE standard distribution systems, such as IEEE 13-bus system and IEEE 123-bus system, only have primary network. (ii) Customers in the real distribution grid model \cite{Realsystem} are equipped with smart meters, which can help us to achieve the proposed online feedback-based linear approximation method. 

\begin{figure}
	\vspace{-0pt} 
	\vspace{-0pt}
	\centering
	\includegraphics[width=1.0\linewidth]{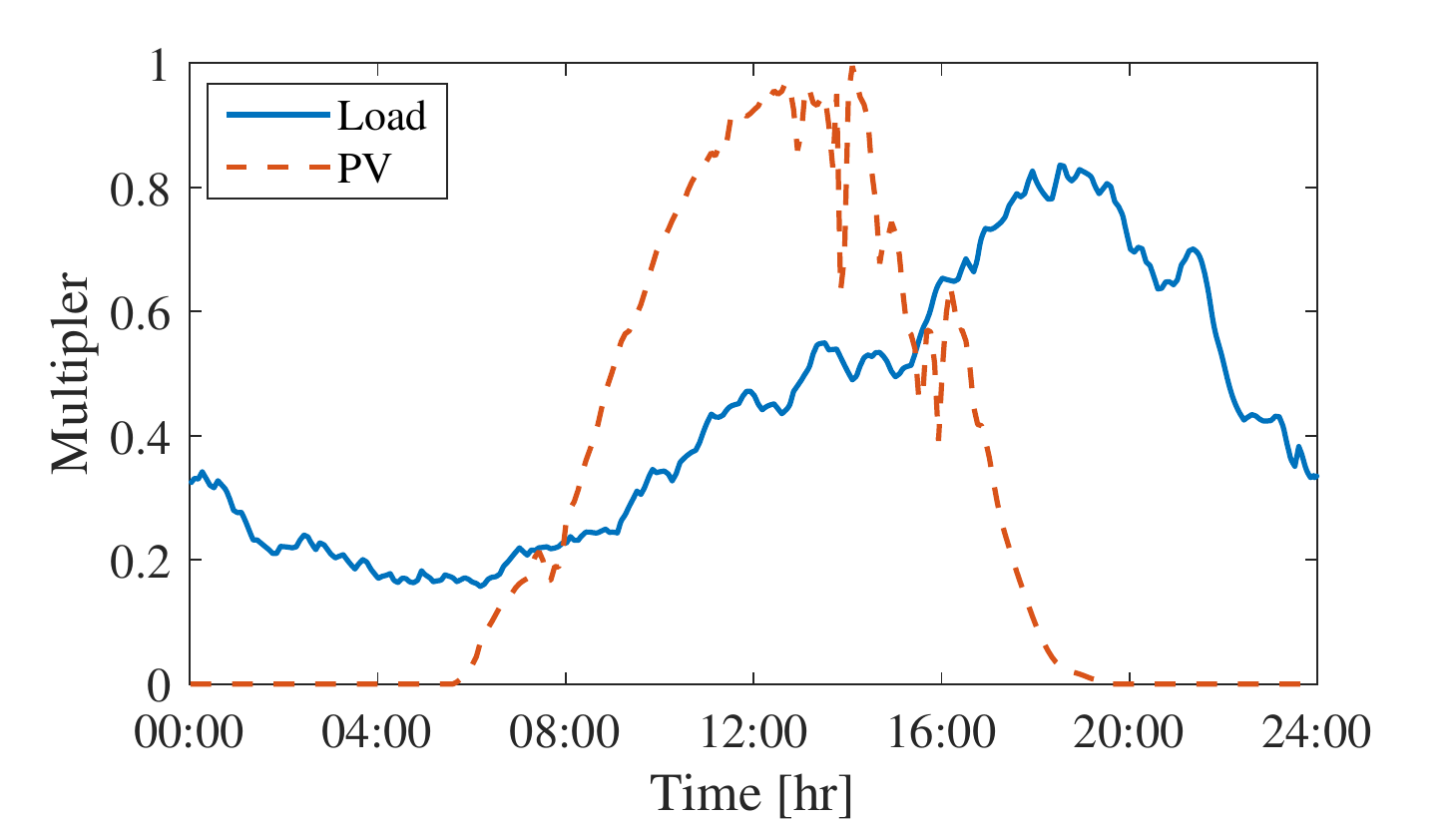}
	\vspace{-0pt} 
	\caption{Time-series multipliers of load demand and PV power.}
	\centering
	\label{load_PV}
    \vspace{-0pt} 
\end{figure}

The time-series multiplier of load demand and solar power with 1-minute time resolution are shown in Fig. \ref{load_PV}. In the case study, PV smart inverters are installed in the secondary networks and the total capacity of PV can serve 30\% load. The base voltages in the primary distribution network and the secondary networks are 13.8 kV and 0.208 kV, respectively. The base power value is 100 kVA. The selected parameters for simulations are summarized in Table \ref{table_Parm}, where the choice of hyper-parameters depends on cross-validation. In general, a bad choice of hyper-parameter will affect the convergence speed and the results. For example, a very large value of the initial penalty factor $\rho$ may lead to a sub-optimal solution, while a too small value of $\rho$ will cause a slow convergence speed. The choice of updating factor $\mu$ has the similar impacts on convergence speed and results. In Table \ref{table_Parm}, the ZIP coefficients of active and reactive loads follow \cite{CVR_PNNL}.

We develop a simulation framework in MATLAB R2019b, which integrates YALMIP Toolbox with IBM ILOG CPLEX 12.9 solver for optimization, and the Open Distribution System Simulator (OpenDSS) for power flow analysis. The OpenDSS can be controlled from MATLAB through a component object model interface, allowing us to carry out the feedback-based linear approximation, performing power flow calculations, and retrieving the feedback results. In this section, we present the convergence analysis to show the impact of asynchronous update on convergence speed. We also demonstrate the effectiveness of our proposed method through numerical evaluations on several benchmarks to study load consumption reduction through CVR implementation: (i) The base case is generated by setting the unity-power factor control mode for all PV inverters where no additional reactive power support is considered. (ii) The VVO-CVR problem is solved by a centralized solver, where the nonlinear terms $\varepsilon_{ij}^{p},\varepsilon_{ij}^{q}$ and $\varepsilon_{ij}^{v}$ in power flow equations are neglected. (iii) The VVO-CVR problem is solved by the proposed distributed method, which requires globally synchronous updates between the leader controller and all the follower controllers. (iv) The VVO-CVR problem is solved by the proposed distributed method with asynchronous updates. The performance testing for different numbers of secondary networks (follower controllers) in the asynchronous distributed algorithm will be presented, where the secondary networks are random selected in each iteration to imitate the possible communication failure or delay in the practical cases. For example, if the number of secondary networks (follower controllers) is set to be 20 in the asynchronous implementation, it will have 20 follower controllers to update and communicate with the leader controller in each iteration. The rest of follower controllers, which are not selected, will remain unchanged in this iteration.

\begin{table}[]
		\centering
		\renewcommand{\arraystretch}{1.3}		
		\caption{Selected parameters}
	    \label{table_Parm}
\begin{tabular}{ccc}
\hline\hline
Description                         & Notion                                & Value            \\ \hline
Initial penalty factor              & $\rho$                                & 0.05                \\ 
Updating factor                     & $\mu$                                 & 10               \\ 
Increasing/Decreasing factor        & $\tau^{\rm inc}$,$\tau^{\rm dec}$     & 5,5                \\ 
Active load ZIP Coefficients        & $k^p_{1},k^p_{2},k^p_{3}$             & 0.96,$-$1.17,1.21  \\
Reactive load ZIP Coefficients      & $k^q_{1},k^q_{2},k^q_{3}$             & 6.28,$-$10.16,4.88  \\
\hline\hline
\end{tabular}
\end{table}

\subsection{Convergence Analysis}\label{sec:convergence}
\begin{figure}
	\vspace{-0pt} 
	\vspace{-0pt}
	\centering
	\includegraphics[width=1.0\linewidth]{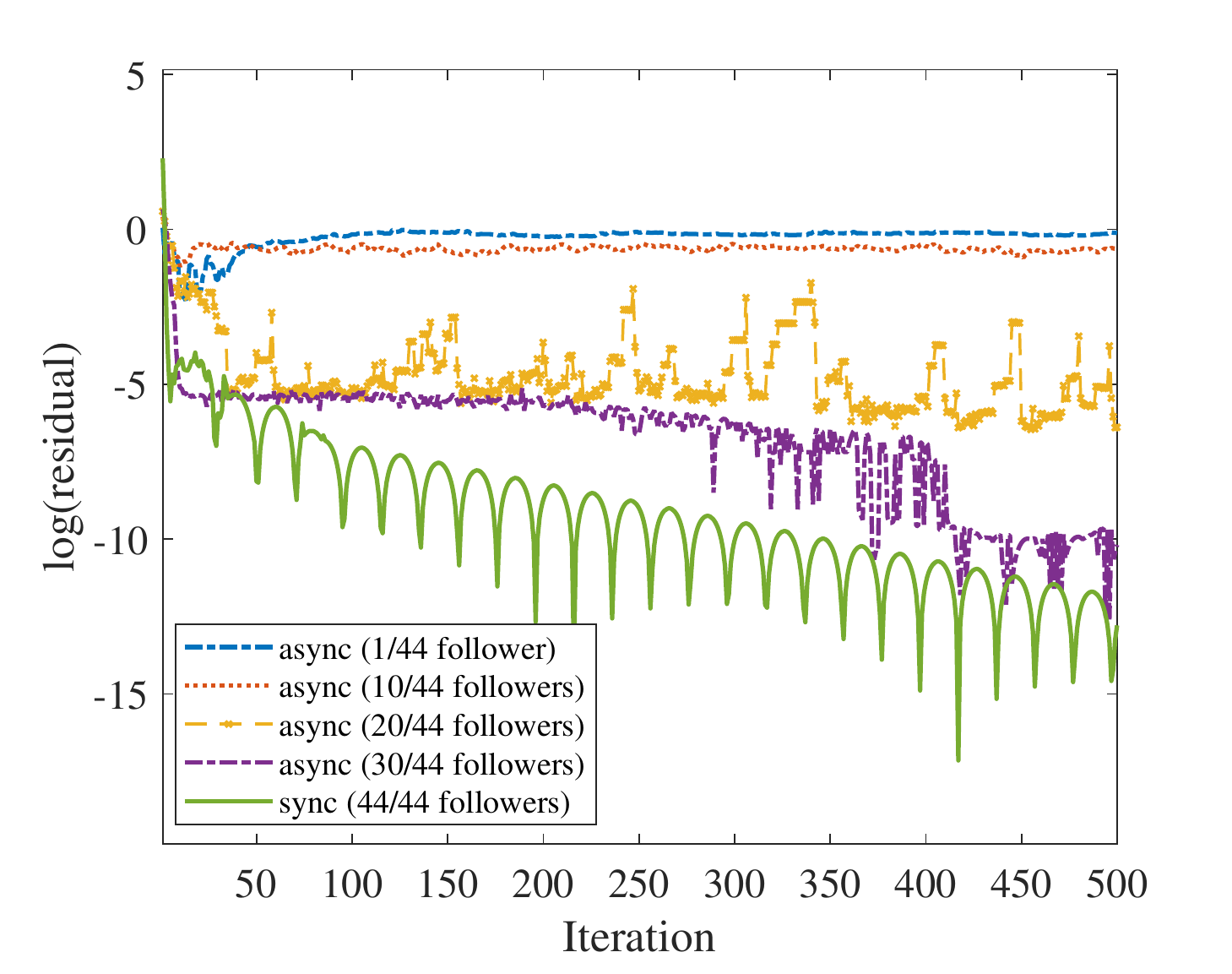}
	\vspace{-0pt} 
	\caption{Convergence speed of the proposed distributed method with synchronous and asynchronous implementation.}
	\centering
	\label{convergence_2}
    \vspace{-0pt} 
\end{figure}

The logarithm values of the norm of primal residuals \eqref{rho_2} with synchronous and different asynchronous communication settings are illustrated in Fig. \ref{convergence_2}, which can be considered as one indicator of the convergence speed for the synchronous and asynchronous updates with different numbers of secondary networks (follower controllers). It can be observed that, if there is no communication failure or delay, the proposed distributed algorithm with the standard ADMM can achieve the best convergence speed; the asynchronous implementation with 20 or 30 activated secondary networks (follower controllers) can still guarantee the convergence with an acceptable speed; while the performance of convergence with 10 or even less secondary networks (follower controllers) are not as good as other cases. Hence, there is a trade-off between the work stress/need on communication system and the convergence performance. The principle of partial barrier is balancing the trade-off between the work stress/need on communication system and the performance of convergence. In our case, the threshold of the number of secondary networks (follower controllers) is 20 to maintain the calculation accuracy. Here, the acceptable speed can be quantified as: if the  primal residuals is lower than $10^{-3}$ within 30 iterations, then we consider the convergence speed is acceptable. Keep in mind that the thresholds may vary in different cases, which should be adjusted accordingly.  

\begin{figure}
	\vspace{-0pt} 
	\vspace{-0pt}
	\centering
	\includegraphics[width=1.0\linewidth]{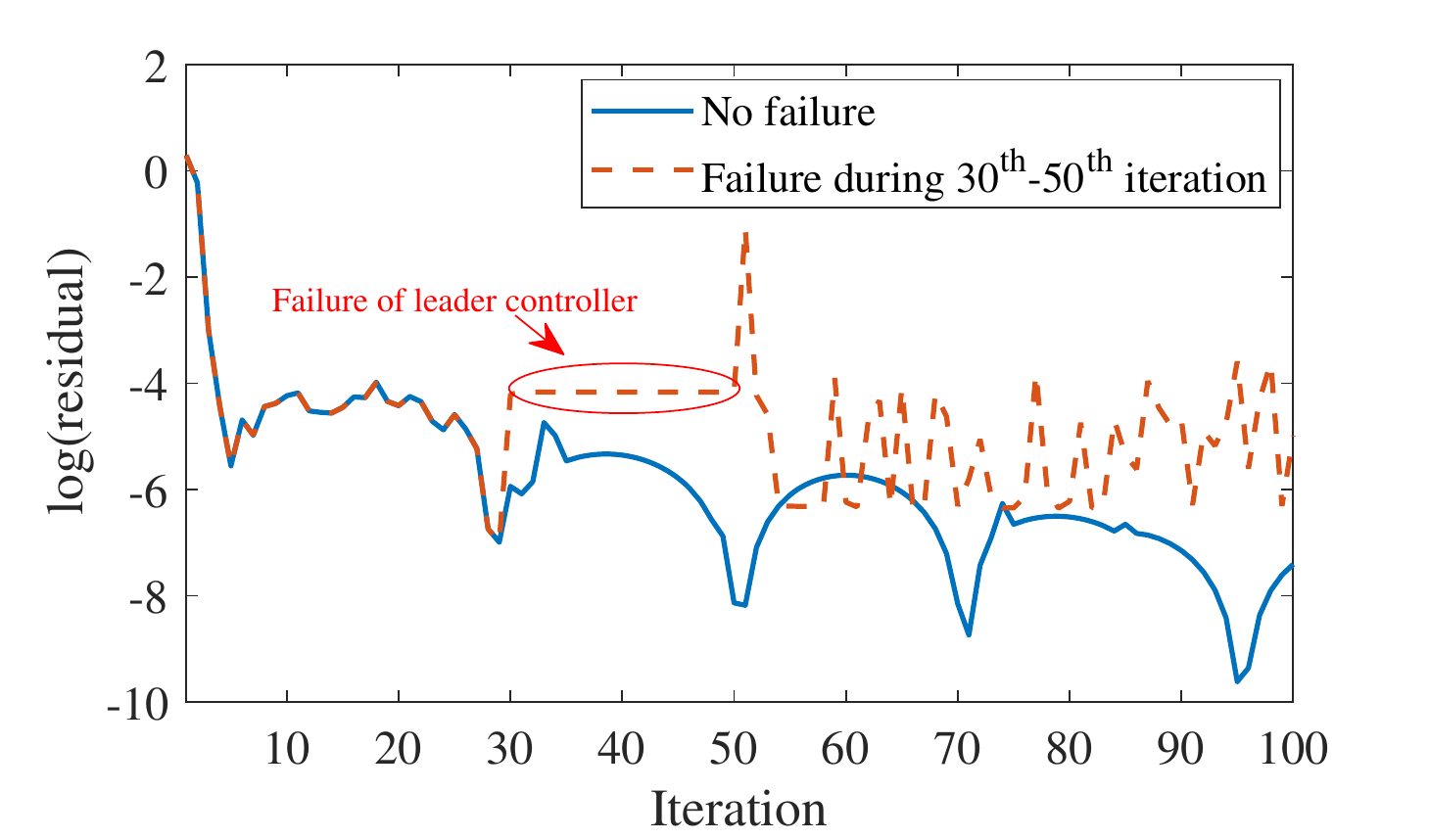}
	\vspace{-0pt} 
	\caption{Convergence speed of the proposed distributed method by considering the potential failure of the primary network (leader controller).}
	\centering
	\label{convergence_3}
    \vspace{-0pt} 
\end{figure}

The distributed leader-follower methods may suffer from the reliability issues when considering the potential failure of the leader controller. To show the impacts of the potential failure of the primary network (leader controller), the convergence speeds of normal communication and communication failure of primary network (leader controller) are compared. In this case, we assume that the primary network (leader controller) could have communication failure by not updating its own sub-problem and communicating with secondary networks (follower controllers) during 30th to 50th iteration, then recover the communication at 51st iteration. In Fig. \ref{convergence_3}, it can be observed that the overall convergence speed is still acceptable even the primary network (leader controller) fails to update and communicate for 20 iterations. Therefore, our proposed distributed algorithm is still efficient for certain level of communication failure of primary network (leader controller).

\subsection{Effect of Online Feedback-Based Approximation}\label{sec:feedback}
\begin{figure}
	\vspace{-0pt} 
	\vspace{-0pt}
	\centering
	\includegraphics[width=1.0\linewidth]{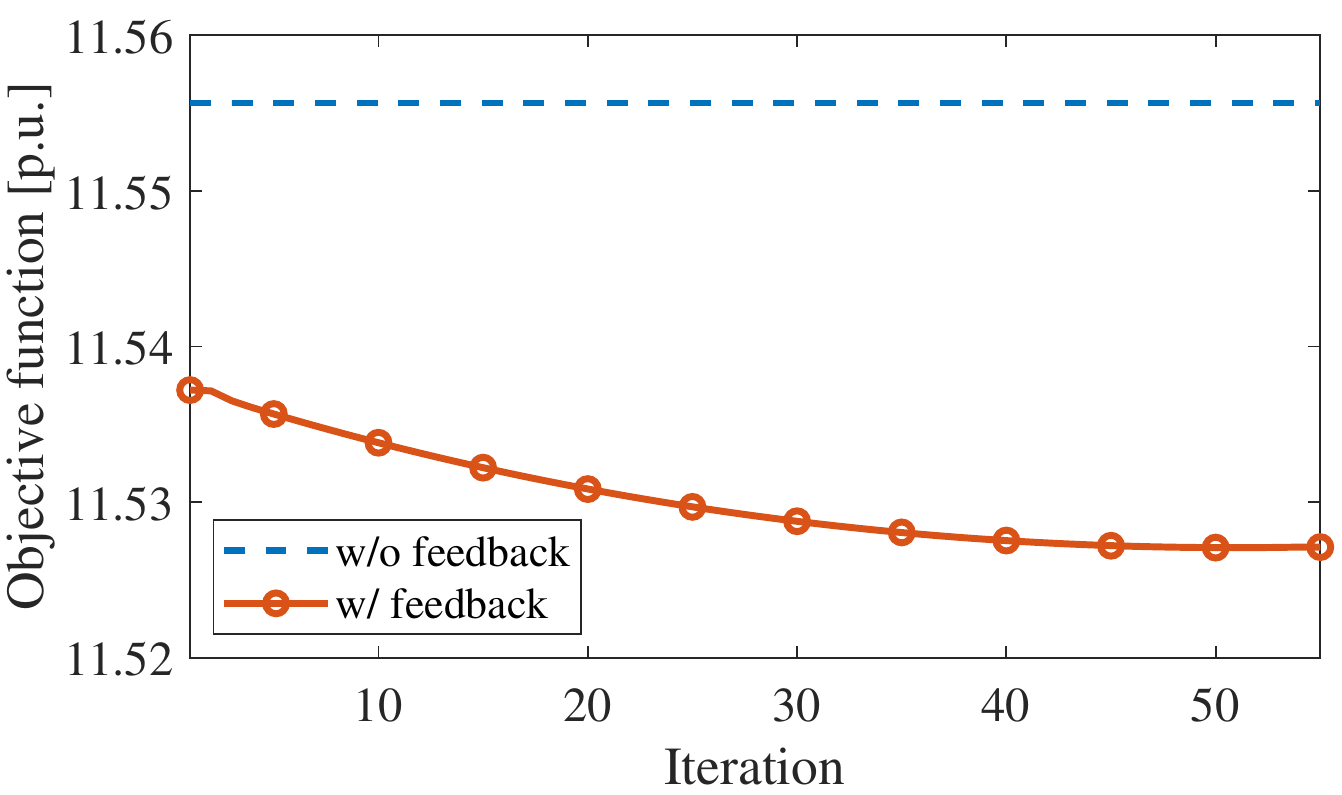}
	\vspace{-0pt} 
	\caption{Objection function values under a fixed-point test.}
	\centering
	\label{convergence_1}
    \vspace{-0pt} 
\end{figure}

To show the effect of online feedback measurements, we solve the VVO-based CVR problem at a fixed point (at 19:00) with different control strategies in centralized and distributed manners. The iterative objective function values (the active power flow through substation) are recorded in Fig. \ref{convergence_1}. Even though the difference of the objective solutions between the centralized solver (blue dashed line) and the proposed distributed method (red line) is about 0.26\% after nearly 50 iteration, the proposed method can still achieve a better result than the centralized method. It is because the proposed distributed method can use measurements feedback from the system to approximate the nonlinear terms successively, while the centralized method neglects the nonlinear terms.

\begin{figure}
	\vspace{-0pt} 
	\vspace{-0pt}
	\centering
	\includegraphics[width=1.0\linewidth]{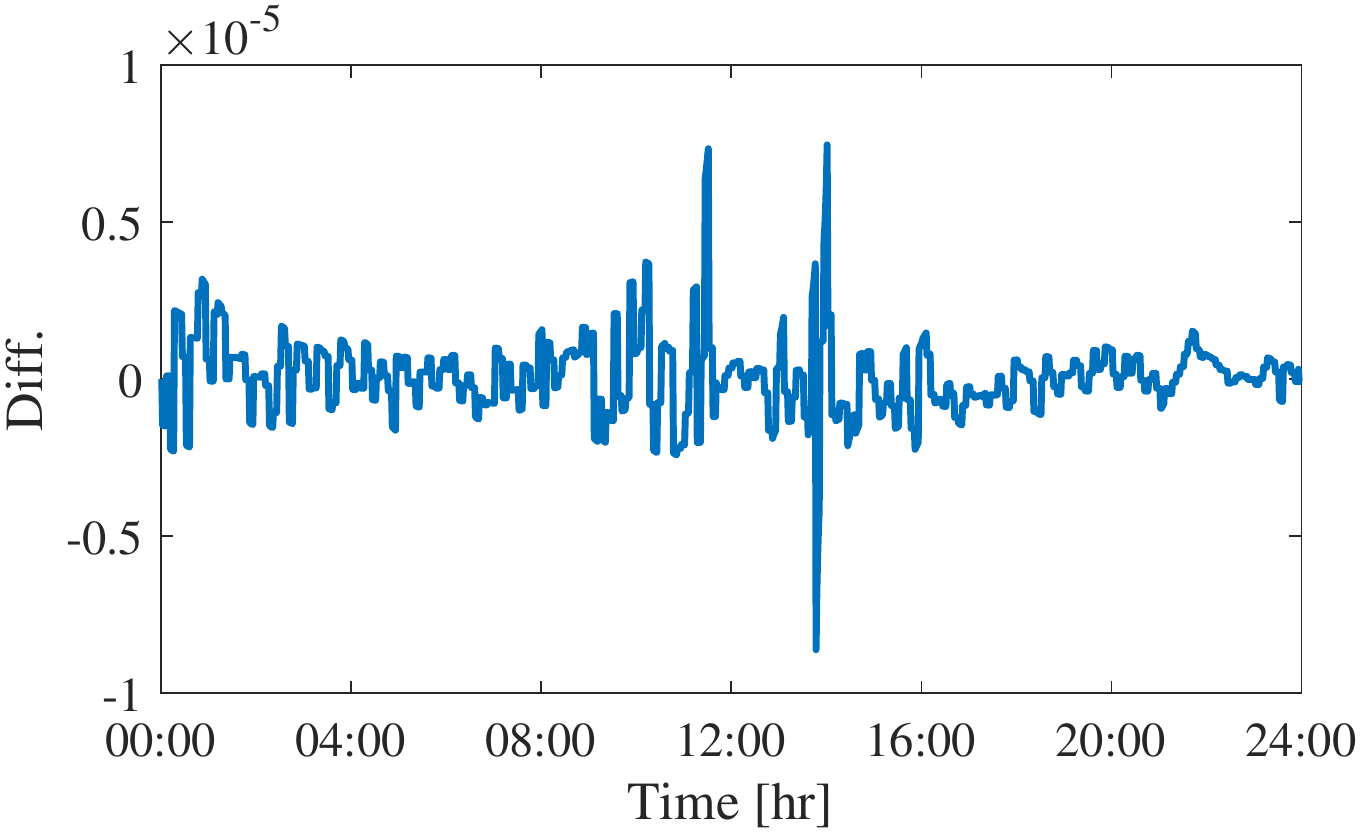}
	\vspace{-0pt} 
	\caption{Difference between the accurate ZIP load and the approximate ZIP load.}
	\centering
	\label{diff_ZIP}
    \vspace{-0pt} 
\end{figure}

To show the effect of approximation of the nonlinear part $\sqrt{v_{i,\phi,t}}$ in (16), we calculate the difference between the accurate ZIP load and the approximate ZIP load with a given time series voltage (1-minute time resolution). The accurate ZIP load at time $t$ is calculated based on the original ZIP load model (4e)--(4f) with the instantaneous voltage at time $t$. While the approximate ZIP load is estimated based on (16)--(18) with the voltage measurement of previous time $t-1$. In Fig. \ref{diff_ZIP}, it can be observed that if the voltage difference between $t$ and $t-1$ is not large, then the differences between the accurate ZIP load and approximate ZIP load are ranging from $-10^{-5}$ to $10^{-5}$, which is acceptable.  

\subsection{Grid-Edge Voltage Profile}\label{sec:secondary}
\begin{figure}
	\vspace{-0pt} 
	\vspace{-0pt}
	\centering
	\includegraphics[width=1.0\linewidth]{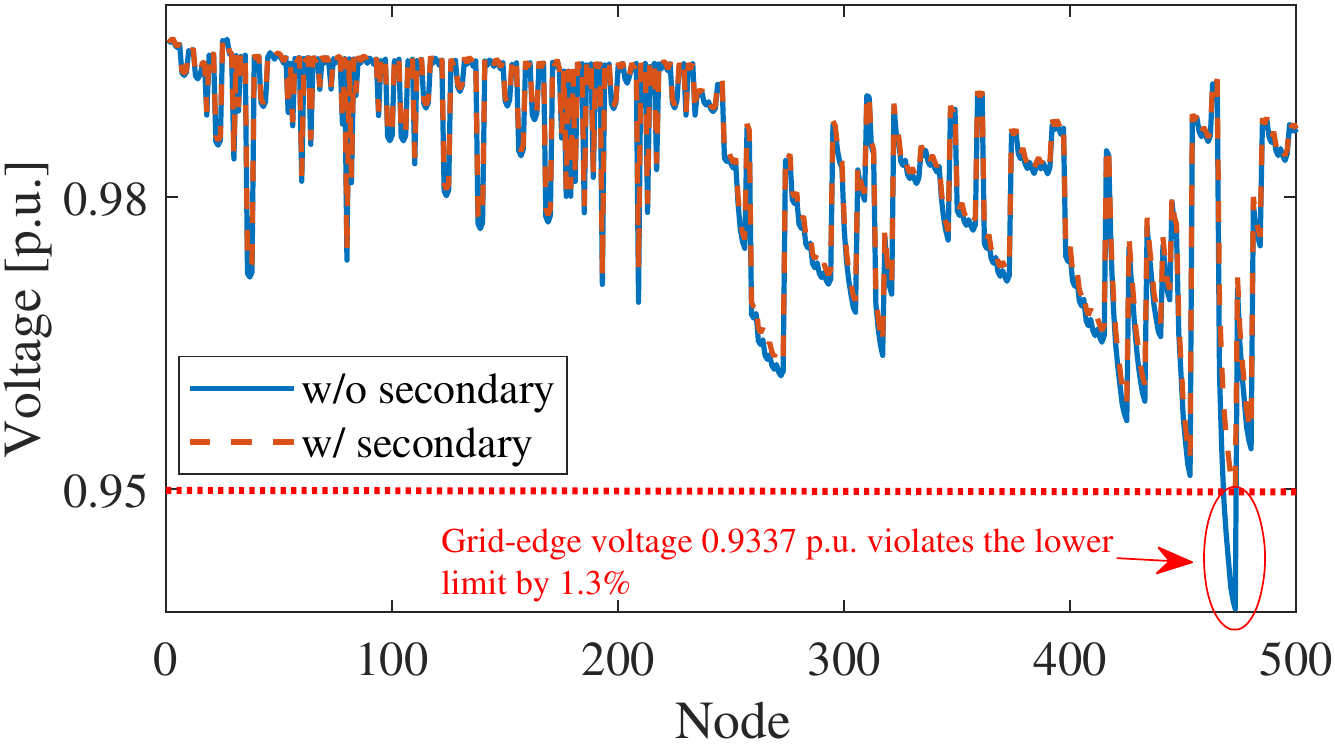}
	\vspace{-0pt} 
	\caption{Nodal voltage profiles with and without the secondary networks.}
	\centering
	\label{V_secondary}
    \vspace{-0pt} 
\end{figure}

In real distribution system, most loads and residential DERs are connected to secondary networks. If the secondary networks are simplified by using aggregate models in primary network, it will hinder the performance of grid-edge voltage regulation. To show the importance of considering detailed models of secondary networks in CVR implementation, two cases are presented: we solve the optimal CVR with and without considering detailed secondary network models, then input the optimal reactive power dispatch results of smart inverters in the distribution system to evaluate the CVR performance. If the secondary networks are not considered in the optimal CVR, the optimal reactive power setting at each primary node has to be proportionally distributed to PV inverters in the secondary networks. The primary and secondary nodal voltage profiles of the two cases are presented in Fig. \ref{V_secondary}, respectively. It can be observed that the grid-edge voltages can be well regulated if both primary and secondary networks are considered in the optimal CVR. However, the grid-edge voltage within one secondary network is 0.9377 p.u., which violates the voltage lower limit 0.95 p.u. by 1.3\%, if we only consider the primary network and aggregate secondary networks as nodal injections.

\subsection{Reactive Power output of Smart Inverters}\label{sec:reactive}
\begin{figure}
	\vspace{-0pt} 
	\vspace{-0pt}
	\centering
	\includegraphics[width=1.0\linewidth]{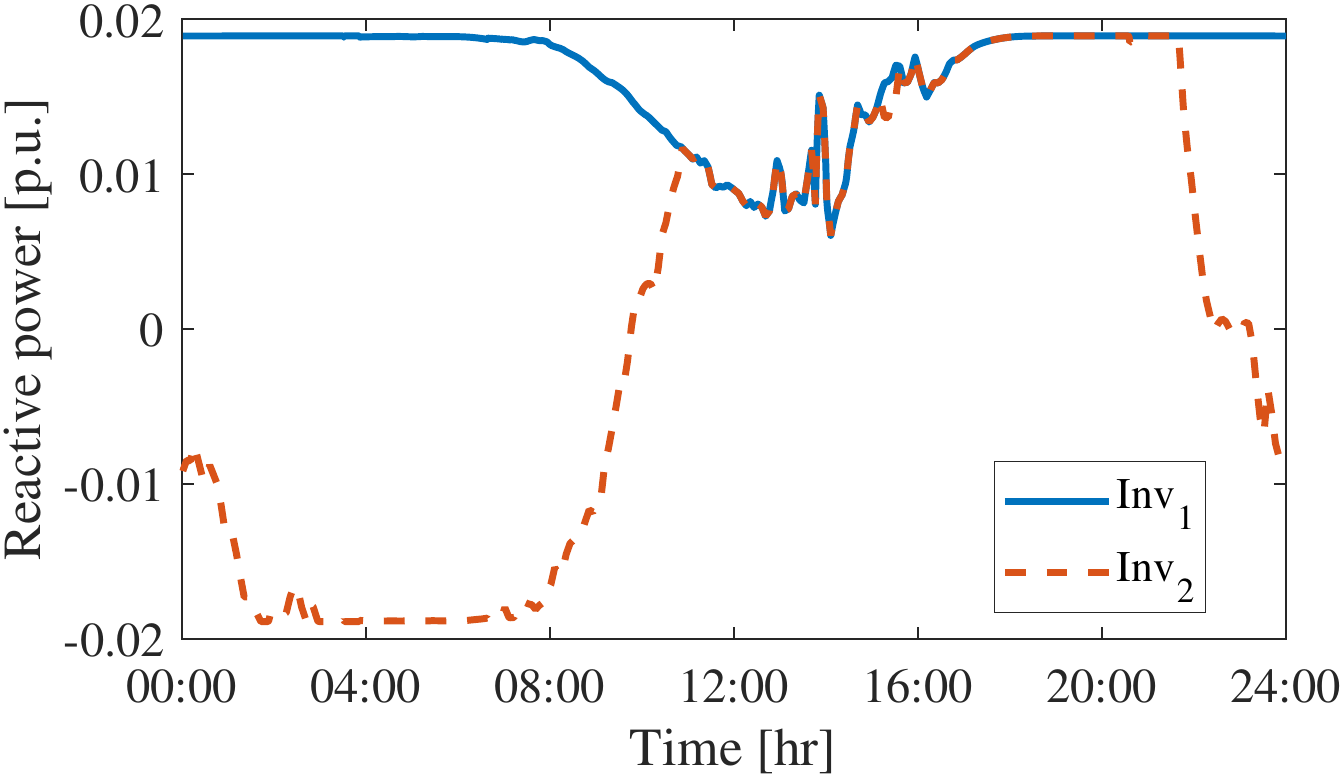}
	\vspace{-0pt} 
	\caption{Reactive power output of two smart inverters as examples.}
	\centering
	\label{reactive}
    \vspace{-0pt} 
\end{figure}

In this test case, there are forty-four secondary networks, and each secondary network are installed with two smart inverters, one in the middle and one in the end of the secondary network. Note that the optimal position and sizing of inverter are not included in the scope of this work. To show the reactive power of inverters in a clear way, we select two inverters as examples with different reactive power behaviors. As shown in Fig. \ref{reactive}, the inverter 1 (blue curve) is installed in the end of the secondary network, where the reactive power injections are always required to maintain the voltage above the lower voltage limit; while the inverter 2 (red dashed curve) is installed in the middle of the secondary network, where the reactive power injection and absorbing are both required to maintain the voltage within predefined voltage limits. Therefore, the reactive power output of inverter will be affected by the installation positions. In our case, it is possible that the reactive power outputs of inverter reach its capacity. For example, because the inverter 1 is installed in the end of a long secondary feeder, our proposed optimal CVR determines inverter 1 to inject enough reactive powers, which satisfy both the reactive power capacity constraints and voltage limit constraints.    

\subsection{Comparison Between Different Control Strategies}\label{sec:timeseries}
To show the time-series simulation, the VVO-CVR is performed in a daily operation of the integrated primary-secondary distribution grid (with 1-minute time resolution) with different control strategies in centralized and distributed manners, respectively. Note that the online implementation of the async-ADMM method is used here, where the nonlinear terms of the network and load models are approximated with the power and voltage measurements feedback from the system with the last-minute dispatch. Existing studies \cite{SM_1,SM_2} have been conducted based on smart meters with 1-minute time resolution. Therefore, the online implementation of the async-ADMM method can by achieved by using 1-minute time resolution measurements sent by smart meters. We also assume that the change of the system is not that large within 1-minute, so that the measurements from the last-minute can still be used to approximate the nonlinear term for the next minute.

\begin{figure}
	\vspace{-0pt} 
	\vspace{-0pt}
	\centering
	\includegraphics[width=1.0\linewidth]{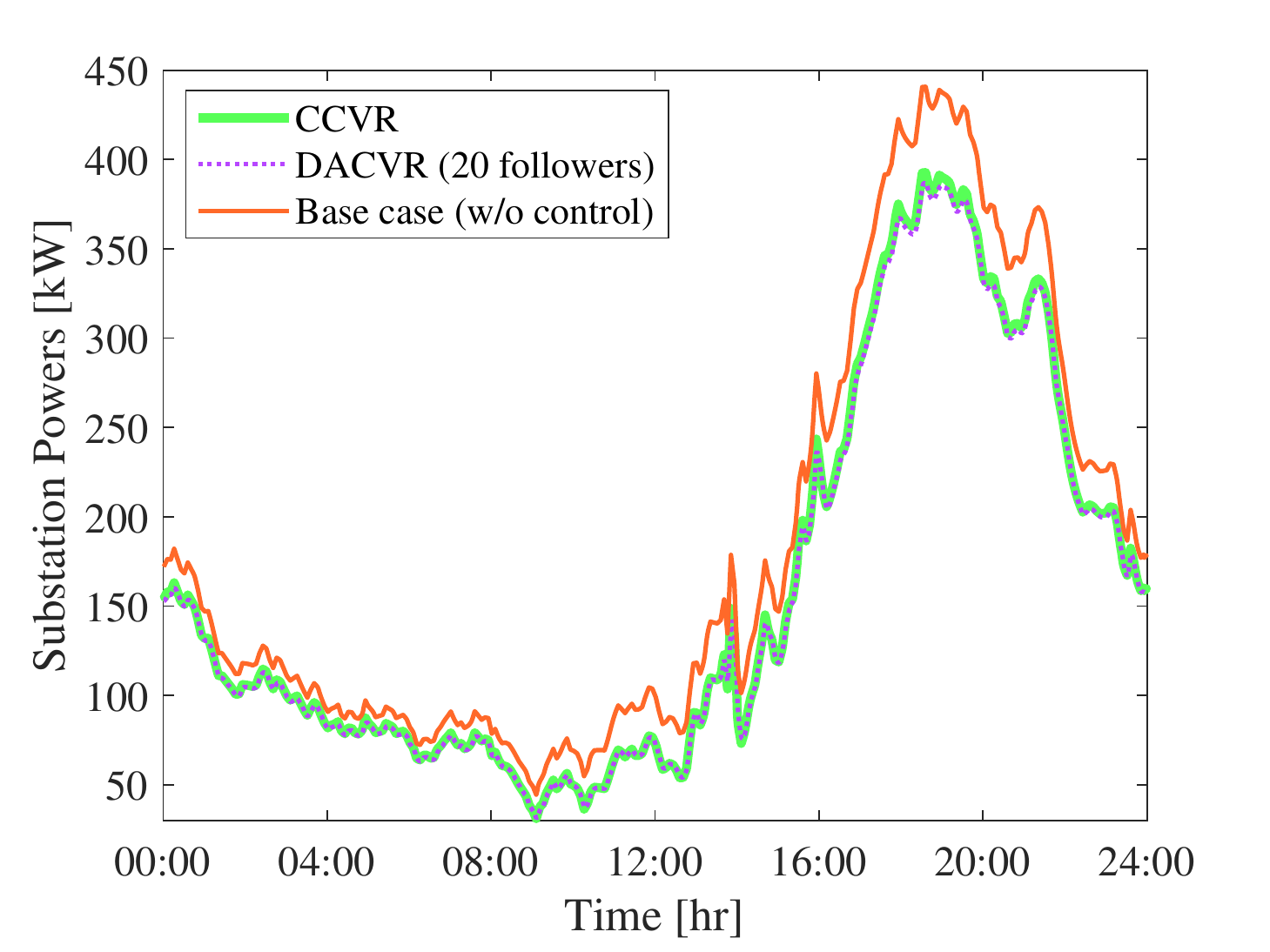}
	\vspace{-0pt} 
	\caption{Substation feed-in active power with different control strategies.}
	\centering
	\label{Psub}
    \vspace{-0pt} 
\end{figure}

The active power supplies from the substation of the base case (without control), centralized CVR (CCVR) and distributed async. CVR (DACVR) with 20 secondary networks (follower controllers) are shown in Fig. \ref{Psub}. As can be observed, the proposed method can effectively reduce the power supply from substation, especially during the peak load period, e.g., 16:00–20:00. To verify the online performance of the proposed distributed method, we compare the time-series solutions of the CCVR (green curve) with DACVR with 20 followers (purple dotted curve). It can be seen that, the DACVR with 20 followers can provide a similar control performance to CCVR. Therefore, when there are at least 20 follower controllers updating and communicating with leader controller in the asynchronous implementation, a good control performance can be achieved.          

The numerical comparisons of total energy consumption over one day and the energy reduction are presented in Table \ref{table_energy_saving} among the base case, CCVR, and distributed sync. CVR (DSCVR) and DACVR with 20 followers. Compared to the base case, the VVO-based CVR method can achieve the energy reduction around 13.2\% to 13.6\%. In theory, the differences between CCVR, DSCVR and DACVR shall be small, because they are solving the similar VVO-CVR problems. The reasons why they do not have the exact same solution are: (i) Because of the missing nonlinear terms in power flow calculations, CCVR cannot obtain the accurate solution; (ii) DACVR obtain the solution by receiving updates from limited number of secondary networks (follower controllers). Based on the comparison between CCVR and DSCVR and DACVR, it can be seen that the total energy consumption results from the CCVR, DSCVR and DACVR are very similar, and DSCVR yields slightly better results than other two cases. This is because DSCVR has the online power and voltage feedback measurements from the system to accurately approximate the nonlinear terms of the power flow calculations and ZIP load models. While the nonlinear terms $\varepsilon_{ij,\phi,t}^{p}, \varepsilon_{ij,\phi,t}^{q}$ and $\varepsilon_{i,\phi,t}^{v}$ are neglected in CCVR, this offline linear approximation method may bring inaccurate power flow and bus voltage computations, consequently, hindering the CVR performance. The energy reduction of DACVR is also slightly less than DSCVR, because DACVR only receives updates from limited number of follower controllers, while DSCVR can receive updates from all follower controllers. It is concluded that DACVR can still obtain a good energy reduction performance with updates from limited number of follower controllers. Compared to CCVR, the advantages of the proposed DSCVR and DACVR can be summarized as follows: (i) The CCVR is disadvantageous on scalability, because CCVR must solve a large-scale VVO-CVR problem. With increasing size of decision models, the computation burden of CCVR increases extensively. While the the proposed DSCVR and DACVR decompose the large-scale problem into multiple small-scale sub problems, therefore, the computation burden is reduced. (ii) In the proposed DSCVR and DACVR, the data privacy and ownership of customers are respected, including local consumption measurement data and cost functions. However, CCVR requires the system-wide collection of data, and a costly communication infrastructure to enable information passing between a control center and regulation devices. (iii) Moreover, the CCVR are susceptible to single point of failure. While DACVR is resilient against agent communication failure or limited communication.   
\begin{table}[]
		\centering
		\renewcommand{\arraystretch}{1.3}		
		\caption{Energy Consumption Results With Different Control Strategies}
		\label{table_energy_saving}
			\vspace{-0pt}\
\begin{tabular}{lcc}
\hline\hline
                                   & Energy (kWh) & Reduction (\%) \\\hline
Base case (w/o control)                        & 262,167.4   &   -             \\
CCVR                 & 227,269.9  & 13.3\%       \\
DSCVR           & 226,339.5  & 13.6\%       \\
DACVR (20 followers)   & 227,325.1  & 13.2\%       \\
\hline\hline
\end{tabular}
\end{table}

\begin{figure}
\centering
\subfloat[Base case (w/o control) \label{voltage_a}]{
\includegraphics[width=1.0\linewidth]{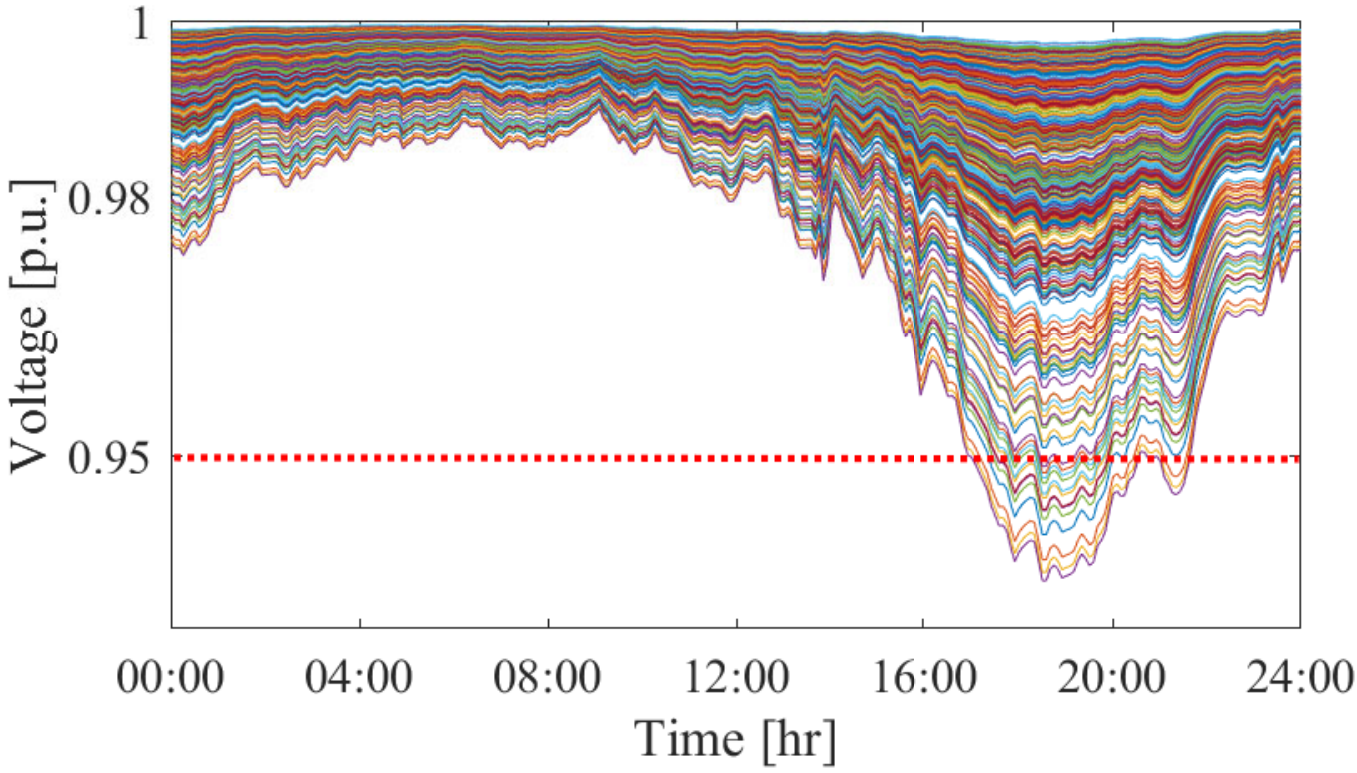}
}
\hfill
\subfloat[DACVR with 20 followers. \label{voltage_b}]{
\includegraphics[width=1.0\linewidth]{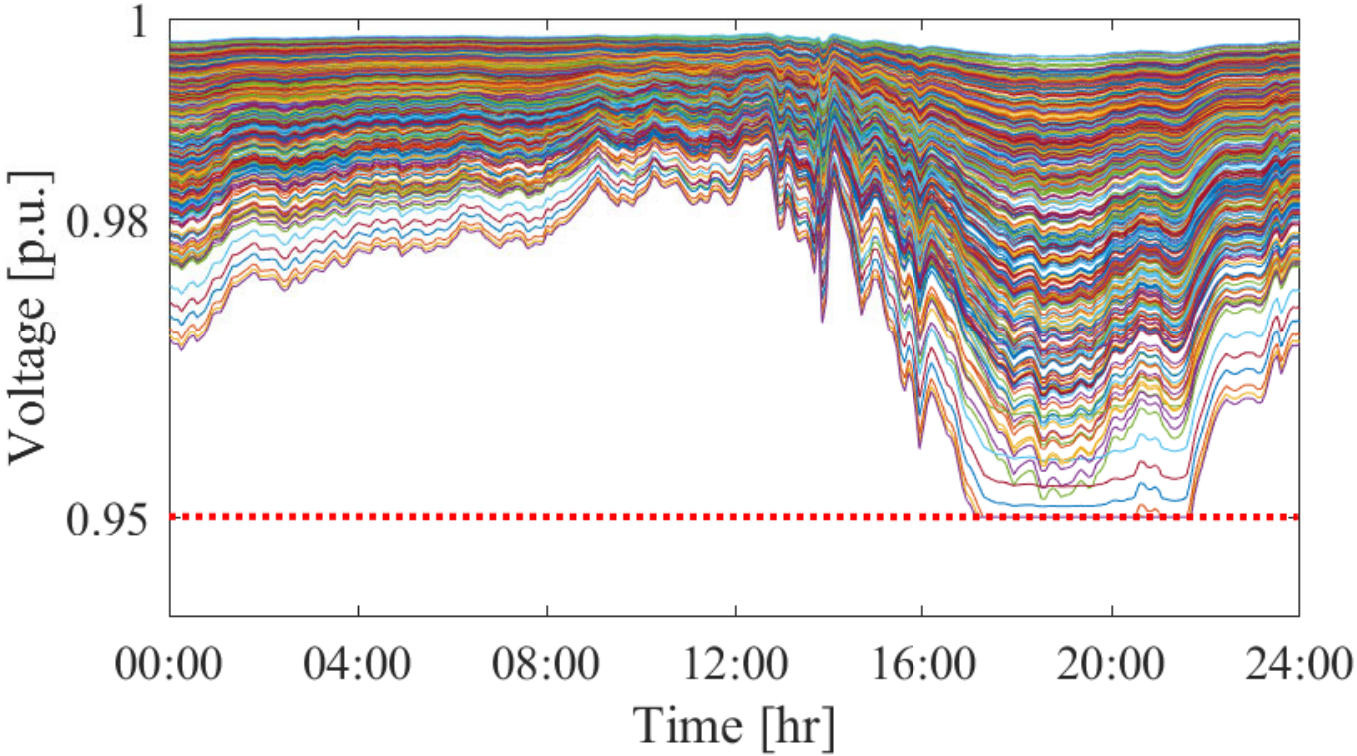}
}
\caption{Voltage profiles with different control strategies (each line represents a phase-wise voltage magnitude of a bus).}
\label{voltages}
		\vspace{-0pt}\
\end{figure}

In Fig. \ref{voltages}, the 1440-minute time-varying voltage profiles of the base case and DACVR with 20 followers are compared. Each line represents a phase-wise voltage magnitude of a bus. As shown in  Fig. \ref{voltages} (a), where there is no reactive power control in the base case, there are voltage violations of the lower limit 0.95 p.u., during the heavy-load periods, e.g., 16:00–20:00. On the other hand, when the CVR is implemented with optimal reactive power control, the system achieves maximum voltage reduction while maintains voltage levels with the predefined range [0.95,1.05] p.u., as shown in  Fig. \ref{voltages} (b).   

\section{Conclusion}\label{sec:Con}
To better regulate voltages at the grid-edge while implementing CVR in distribution system, a distributed VVO-CVR algorithm is developed to optimally coordinate the smart inverters in unbalanced three-phase integrated primary-secondary distribution systems. In order to handle the non-convexity of power flow and ZIP load models, a feedback-based linear approximation method has been proposed to successively estimate the nonlinear terms in these models. An ADMM-based distributed framework is established to solve the optimal CVR problem in a leader-follower distributed fashion, where the primary system corresponds to the leader controller and each secondary system corresponds to a follower controller. We further address its asynchronous implementation with a frozen strategy that allows asynchronous updates. Simulation results on a real Midwest U.S. distribution feeder have validated the robustness and effectiveness of the proposed method. According to the case studies, we have shown that: (1) With a reasonable setting of asynchronous update, the proposed async-ADMM method is able to guarantee the convergence with acceptable speed. (2) Compared to using aggregate models of secondary networks, the grid-edge voltages can be better regulated with detailed secondary network models in the proposed CVR implementation. (3) With the online feedback-based linear approximation, the proposed VVO-CVR can achieve good performance of energy/voltage reductions while maintaining voltage level in predefined ranges.

% or


% % you can choose not to have a title for an appendix
% % if you want by leaving the argument blank
% \section{}



\ifCLASSOPTIONcaptionsoff
  \newpage
\fi



\bibliographystyle{IEEEtran}
\bibliography{IEEEabrv,./bibtex/bib/IEEEexample}

\begin{IEEEbiography}[{\includegraphics[width=1in,height=1.25in,clip,keepaspectratio]{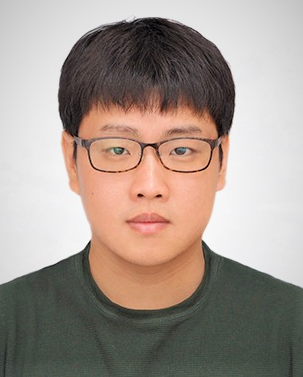}}]{Qianzhi Zhang}(S'17) is currently pursuing his Ph.D. in the Department of Electrical and Computer Engineering, Iowa State University, Ames, IA. He received his M.S. in electrical and computer engineering from Arizona State University in 2015. From 2015 to 2016, he worked as a research engineer with Huadian Electric Power Research Institute. His research interests include the applications of machine learning and advanced optimization techniques in power system operation and control.
\end{IEEEbiography}
	
\begin{IEEEbiography}[{\includegraphics[width=1in,height=1.25in,clip,keepaspectratio]{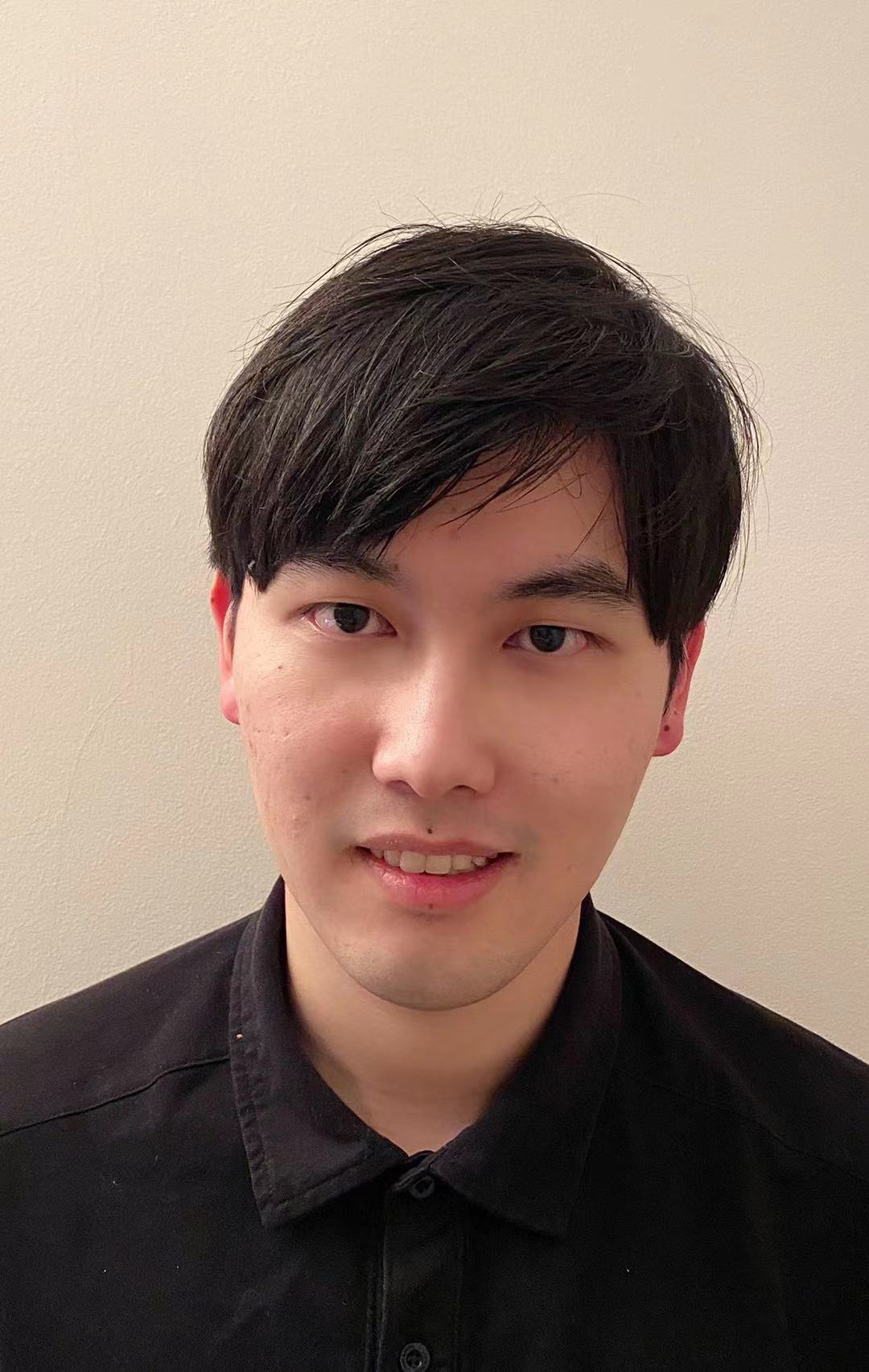}}]{Yifei Guo}(M'19) received the B.E. and Ph. D. degrees in electrical engineering from Shandong University, Jinan, China, in 2014 and 2019, respectively.
Currently, he is a Postdoctoral Research Associate with the Department of Electrical and Computer Engineering,  Iowa State University, Ames, IA, USA. He was a visiting student with the Department of Electrical Engineering, Technical University of Denmark, Lyngby, Denmark, in 2017--2018. 

His research interests include voltage/var control, renewable energy integration, wind farm control, distribution system optimization and control, and power system protection.
\end{IEEEbiography}

\begin{IEEEbiography}[{\includegraphics[width=1in,height=1.25in,clip,keepaspectratio]{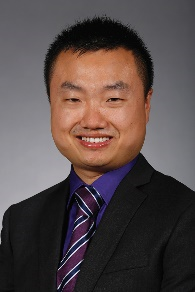}}]{Zhaoyu Wang}(S'13--M'15--SM’20) is the Harpole-Pentair Assistant Professor with Iowa State University. He received the B.S. and M.S. degrees in electrical engineering from Shanghai Jiaotong University, and the M.S. and Ph.D. degrees in electrical and computer engineering from Georgia Institute of Technology. His research interests include optimization and data analytics in power distribution systems and microgrids. He is the Principal Investigator for a multitude of projects focused on these topics and funded by the National Science Foundation, the Department of Energy, National Laboratories, PSERC, and Iowa Economic Development Authority. Dr. Wang is the Chair of IEEE Power and Energy Society (PES) PSOPE Award Subcommittee, Co-Vice Chair of PES Distribution System Operation and Planning Subcommittee, and Vice Chair of PES Task Force on Advances in Natural Disaster Mitigation Methods. He is an editor of IEEE Transactions on Power Systems, IEEE Transactions on Smart Grid, IEEE Open Access Journal of Power and Energy, IEEE Power Engineering Letters, and IET Smart Grid. Dr. Wang was the recipient of the National Science Foundation (NSF) CAREER Award, the IEEE PES Outstanding Young Engineer Award, and the Harpole-Pentair Young Faculty Award Endowment.
\end{IEEEbiography}

\begin{IEEEbiography}[{\includegraphics[width=1in,height=1.25in,clip,keepaspectratio]{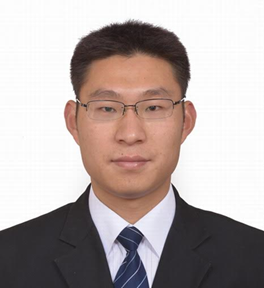}}]{Fankun Bu}(S’18) received the B.S. and M.S. degrees from North China Electric Power University, Baoding, China, in 2008 and 2013, respectively. From 2008 to 2010, he worked as a commissioning engineer for NARI Technology Co., Ltd., Nanjing, China. From 2013 to 2017, he worked as an electrical engineer for State Grid Corporation of China at Jiangsu, Nanjing, China. He is currently pursuing his Ph.D. in the Department of Electrical and Computer Engineering, Iowa State University, Ames, IA. His research interests include distribution system modeling, smart meter data analytics, renewable energy integration, and power system relaying.
\end{IEEEbiography}
\end{document}